\documentclass[aps,pra,twocolumn,showpacs,floatfix]{revtex4-1}
\usepackage{bm}
\usepackage{graphicx} % for PDF output
\usepackage{amsmath}
\usepackage{epstopdf}
\usepackage{subfigure}
\usepackage{array}
\begin{document}
\title{Hyperspherical-LOCV Approximation to Resonant BEC}

\author{M. W. C. Sze$^1$, A. G. Sykes$^2$, D. Blume$^3$, J. L. Bohn$^1$}
\affiliation{$^1$JILA and Department of Physics, University of Colorado, Boulder, Colorado 80309-0440, USA}
\affiliation{$^2$Transpower New Zealand Ltd, Waikoukou, 22 Boulcott Street, PO Box 1021, Wellington}
\affiliation{$^3$Homer L. Dodge Department of Physics and Astronomy,
  The University of Oklahoma, 440 W. Brooks Street, Norman, Oklahoma 73019, USA}

\date{\today}

\begin{abstract}
We study the ground state properties of a system of $N$ harmonically trapped bosons of mass $m$ interacting with two-body contact interactions, from small to large scattering lengths.  This is accomplished in a hyperspherical coordinate system that is flexible enough to describe both the overall scale  of the gas and two-body correlations.  By adapting the lowest-order constrained variational (LOCV) method, we are able to semi-quantitatively attain Bose-Einstein condensate ground state energies even for gases with infinite scattering length.   In the large particle number limit, our method provides analytical estimates for the energy per particle $E_0/N \approx 2.5 N^{1/3} \hbar \omega$ and two-body contact $C_2/N \approx 16 N^{1/6}\sqrt{m\omega/\hbar}$ for a Bose gas on resonance, where $\omega$ is the trap frequency. 

\end{abstract}

%\pacs{}

\maketitle

%%%%%%%%%%%%%%%%%%%%%%%%%%%%%%%%%

\section{Introduction}

Recent experimental progress has pushed the physics of dilute Bose-Einstein condensates (BEC's) into the regime of formally infinite two-body scattering length $a$, by means of tuning magnetic fields on resonance with Fano-Feshbach resonances.  Such gases, termed ``resonant BEC's,'' tend to be short-lived in this limit owing to three-body recombination.  On their way to this catastrophe, however, they nevertheless show signs of quasi-equilibration \cite{Makotyn14_NP} as well as the formation of bound dimer and trimer states of the atoms \cite{Klauss17_PRL, Eigen17_PRL, Fletcher13_PRL, Fletcher17_Science}.  

Experiments in this limit pose problems for ordinary, perturbative field theory, with the product $na^3$ as its small parameter, where $n$ is the number density .  This parameter is obviously no longer small in the resonant limit.  Field-theoretic and other familiar formalisms can be rescued, however, by using renormalized interactions based on finite effective scattering lengths that replace the bare $a$ - which may or may not necessitate an introduction of momentum cutoffs- or else on two-body interactions such as square wells \cite{Song09_PRL,Yin13_PRA,Sykes14_PRA,Yin16_PRA,Corson15_PRA,Corson16_PRA,Ding17_PRA,vanHeugten}.

% by using renormalized interactions based on effective scattering lengths subject to momentum cutoffs, or else on two-body interactions such as square wells \cite{Song09_PRL,Yin13_PRA,Sykes14_PRA,Yin16_PRA,Corson15_PRA,Corson16_PRA,Ding17_PRA}.  

An interesting alternative to these methods is to seek explicit wave function solutions for the resonant BEC.  A particular approach, to be taken in this paper, begins from a hyperspherical coordinate method, which emphasizes collective, rather than independent-particle, coordinates \cite{Bohn98_PRA,Ding17_PRA}.  Roughly, these methods employ a single macroscopic coordinate - the hyperradius -  to denote the collective motion of the condensate as a whole.  When suitably complemented by two-body interparticle coordinates, the method has been extended to include realistic two-body potentials between the atoms \cite{Das04_PRA,Das07_PRA,Chakrabarti08_PRA,Lekala14_PRA}, or else boundary conditions employing realistic (i.e., not renormalized) scattering lengths \cite{Sorensen02_PRA,Sorensen03_PRA,Sorensen04_JPB,Sogo05_JPB}.  

In this article we modify the idea of hyperspherical methods as applied to the trapped, resonant Bose gas.  We arrive at approximate BEC ground state energies and wave functions for a range of scattering 
%DB length 
lengths 
from $a=0$ to $a = \infty$, and for negative $a$ until the collapse threshold is reached.  The key to  this extension is an alternate renormalization procedure, borrowed from quantum Monte Carlo calculations, that is used to generate correlated wave functions.  Specifically, we expand the hyperangular part of the wave function into a Jastrow product wave function \cite{Jastrow55_PR}.  This allows the easy approximate evaluation of the relevant integrals in the theory, and provides a consistent expansion of the energy into the relevant cluster expansion.  

One clear simplification of the theory arises in the use of the lowest-order constrained variational (LOCV) approximation to the integrals \cite{Pandharipande73_PRC,Pandharipande77_PRA}, which has been employed previously for homogeneous Bose gases \cite{Cowell02_PRL}.  Here we meld this approximation with the hyperspherical approach, which results in meaningful approximations to the ground state of a trapped gas, termed the hyperspherical LOCV (H-LOCV) approximation, while also setting the stage for describing an excited state spectrum which will ultimately allow for the calculation of detailed many-body dynamics.  

This paper is organized as follows: Section II discusses the hyperspherical coordinates, and the $N-$body Hamiltonian and wave function expressed in this coordinate system. We introduce the Jastrow wave functions to be used as basis functions and the LOCV method in Section III. Results and their interpretations are presented in Section IV. We draw our conclusions and outlook for future work in Section V.

%%%%%%%%%%%%%%%%%%%%%%%%%%%%

\section{Formalism: General Aspects}

We are interested in solving the Schr\"odinger equation for the $N$-particle system with Hamiltonian
\begin{eqnarray}
H = \sum_{i=1}^N \left( \frac{ p_i^2 }{ 2m } + \frac{ 1 }{ 2 } m \omega^2 r_i^2 \right) + \sum_{i<j}^{N} V({\bf r}_{ij}),
\end{eqnarray}
for a collection of identical bosons of mass $m$ interacting via pairwise potentials $V$, and confined to a harmonic oscillator trap with angular frequency $\omega$.  A typical experimental realization of an ultracold Bose gas is dilute, meaning that the range of the potential $V$ is far smaller than the mean spacing between atoms.  Under these circumstances, the atoms may be considered as mostly independent particles, with Hamiltonian
\begin{eqnarray}
H = \sum_{i=1}^{N} \left( \frac{ p_i^2 }{ 2m } + \frac{ 1 }{ 2 } m \omega^2 r_i^2 \right),
\label{eq:Hamiltonian_V0}
\end{eqnarray}
{\it but} with severe boundary conditions on the many-body wave function whenever two atoms are near one another.  These are referred to as the Bethe-Peierls boundary conditions.  If the total wave function is denoted by $\Psi$, then we require
\begin{eqnarray}
\lim_{r_{ij} \rightarrow 0 } \frac{ 1 }{ (r_{ij} \Psi) } \frac{ \partial (r_{ij} \Psi) }{ \partial r_{ij} } = - \frac{ 1 }{ a }\label{eq:BP_boundary_conditions}
\end{eqnarray}
for any 
%DB pair 
pair distance
$r_{ij}$, where $a$ is the two-body scattering length of the two-body interaction.  More generally, three-body boundary conditions are also required to describe the gas, but we do not consider this for the present.

\subsection{Hyperspherical Coordinates}

Our general strategy is to carefully choose a small set of relevant coordinates for the atoms.  This will include an overall coordinate, the hyperradius, giving the size of the condensate and describing its collective breathing modes; and relative coordinates between pairs of atoms, allowing us to implement the boundary condition (\ref{eq:BP_boundary_conditions}) as well as account for excitations and correlations of atom pairs.  

Specifically, we follow the coordinate system and notation of \cite{Sorensen02_PRA}, defining first  the center of mass coordinate 
\begin{eqnarray}
{\bf R}_{\rm cm} = \frac{ 1 }{ N } \sum_{i=1}^N {\bf r}_i.
\end{eqnarray}
The remaining coordinates are given as a set of $N-1$ Jacobi vectors that locate each atom $k$ from the center of mass of the preceding $k-1$ atoms:
\begin{eqnarray}
{\bm \eta}_k = \sqrt{ \frac{ N-k }{ N-k+1 } } 
\left( {\bf r}_{N-k+1} - \frac{ 1 }{ N-k } \sum_{j=1}^{N-k} {\bf r}_j \right).
\label{eq:eta_k}
\end{eqnarray}
These Jacobi vectors form Cartesian coordinates in a configuration space of dimension $d \equiv 3(N-1)$ denoting the relative motion of the atoms.    The collective coordinate is the hyperradius $\rho$, the radial coordinate in this space, given by
\begin{eqnarray}
\rho^2 =  \sum_{k=1}^{N-1} {\bf \eta}_k^2  =  \frac{ 1 }{ N } \sum_{i<j}^{N} r_{ij}^2 .
\end{eqnarray}
Thus the hyperradius also denotes the root-mean-squared interparticle spacing for any configuration of atoms, a measure of extent of the gas.

The angular coordinates on this hypersphere, collectively denoted by $\Omega$, may be chosen in a great many different ways \cite{Smirnov77,Avery97_IJQC}.  A main point, however, is that all such angular coordinates are bounded and therefore eigenstates of kinetic energy operators expressed in these coordinates have discrete spectra and are characterized by a collection of as many as $3N-4$ discrete quantum numbers.  These are complemented by a description of the motion in $\rho$, which is also bounded within a finite range due to confinement by the harmonic oscillator potential.  The 
%DB wave 
relative wave 
function of the gas can therefore be expanded in a discrete basis set, whose quantum numbers describe the modes contributing to an energy eigenstate, or else the modes excited in a dynamical time evolution of the gas. 

For our purposes we again follow \cite{Sorensen02_PRA}. The full set of $3N-4$ hyperangles $\Omega$ will be partitioned into specific coordinates as follows.  $2(N-1)$ of the hyperangles will simply be the angular coordinates (in real space) of the Jacobi vectors, $(\theta_k, \varphi_k)$. The remaining hyperangles $\alpha_k$ are angles of radial correlation among the Jacobi coordinates,
\begin{eqnarray}
\frac{ \eta_k }{ \left(\sum_{l=1}^k \eta_l^2 \right)^{1/2}} = \sin \alpha_k, \qquad k=2,3,..., N-1.
\end{eqnarray}
(Notice that in this definition $\alpha_1=\pi/2$ is not a separate coordinate \cite{Sorensen02_PRA}.) Among these $\alpha_k$, we single out  the angle that parametrizes a single pair,
\begin{eqnarray}
\sin \alpha \equiv \sin \alpha_{N-1} = \frac{ r_{12} }{ \sqrt{2} \rho }.
\end{eqnarray}
The hyperangle $\alpha$ ranges from 0 (where particles 1 and 2 precisely coincide) to $\pi/2$ (where many-body effects prevail). For our present purposes, we will assume that the orbital angular momentum in each relative coordinate  is zero, whereby the angles $(\theta_k,\varphi_k)$ are irrelevant.  We are also going to restrict our wave functions to those with an explicit dependence on $r_{12}$ only, with the understanding that we must symmetrize the wave function among all pairs $r_{ij}$.  With this assertion,  the wave function calculations will be carried out in the coordinates $(\rho,\alpha)$, the 
%DB ``most relevant'' two 
two ``most relevant'' 
degrees of freedom. 

In this coordinate system, the volume element (to be used later in calculating matrix elements in the basis set) is \cite{Sorensen02_PRA,Aquilanti86_JCP}
\begin{eqnarray}
\rho^{3N-4} d \rho \; d \Omega_{N-1},
\end{eqnarray}
where the surface area element on the hypersphere is defined recursively as
% DB please check my changes here: I inserted a, what I thought is a, missing \alpha_k and I placed the d\alpha in a different position... 
%\begin{eqnarray}
%d\Omega_k = \sin \theta_k d \theta_k \;\; d \varphi_k \; d \alpha_k \sin^2 \alpha_k \cos ^{3k-4} 
%\;\; d\Omega_{k-1}.
%\end{eqnarray}
\begin{eqnarray}
d\Omega_k = \sin \theta_k d \theta_k \;\; d \varphi_k \; 
\sin^2 \alpha_k \cos ^{3k-4} \alpha_k \; d \alpha_k 
\;\; d\Omega_{k-1}.
\end{eqnarray}
The most important volume element for our purposes is that for the hyperangle $\alpha = \alpha_{N-1}$, which we express in the specialized notation
%DB again, please check my change here: I placed the d \alpha in a different position...
%\begin{equation}
%d\Omega_{\alpha} = d \alpha \sin^2 \alpha \cos^{3N-7} \alpha = J(\alpha) d \alpha 
%\end{equation}
\begin{equation}
d\Omega_{\alpha} = 
\sin^2 \alpha \cos^{3N-7} \alpha \; d \alpha = J(\alpha) d \alpha 
\end{equation}
that defines a shortcut notation for the Jacobian $J(\alpha)$.  We also single out the angular coordinates $(\theta,\varphi)$ $=(\theta_{N-1},\varphi_{N-1})$ of this Jacobi 
%DB vector. Thus
vector,
\begin{eqnarray}
d \Omega_{N-1} = \sin \theta d \theta \; d\varphi \; J(\alpha) d\alpha \; d \Omega_{N-2}.
\end{eqnarray}

\subsection{Hamiltonian and Wave Function in Hyperspherical Coordinates}

In the coordinates above, and disregarding the trivial motion of the center of mass, the relative motion component of the Hamiltonian (\ref{eq:Hamiltonian_V0}) reads \cite{Sorensen02_PRA}
\begin{align}
H_{\rm{rel}} = - \frac{ \hbar^2 }{ 2m } \left[ \frac{ 1 }{ \rho^{3N-4} } \frac{ \partial }{ \partial \rho } \rho^{3N-4} 
\frac{ \partial }{ \partial \rho } - \frac{ {\Lambda}_{N-1}^2 }{ \rho^2 } \right] + \frac{ 1 }{ 2 } m \omega^2 \rho^2.
\label{eq:full_H_hyper}
\end{align}
Thus the kinetic energy has a radial part and an angular part, given in general by the grand angular momentum $\Lambda_{N-1}^2$.  Like the surface area element, this angular operator can be defined recursively,
\begin{eqnarray}
\label{eq:Lambda_k}
\Lambda_k^2 = \Pi_k^2 + \frac{ \Lambda_{k-1}^2 }{ \cos^2 \alpha_k } + \frac{ l_k^2 }{ \sin^2 \alpha_k }
%DB .
\end{eqnarray}
with
%DB I placed brackets around the aguments 2 alpha for easier readibility 
%\begin{eqnarray}
%\Pi_k^2 = - \frac{ \partial ^2 }{ \partial \alpha_k^2 } 
%+ \frac{ (3k-6) - (3k-2) \cos 2 \alpha_k }{ \sin 2 \alpha_k } \frac{ \partial }{ \partial \alpha_k }.
%\end{eqnarray}
\begin{eqnarray}
\Pi_k^2 = - \frac{ \partial ^2 }{ \partial \alpha_k^2 } 
+ \frac{ (3k-6) - (3k-2) \cos (2 \alpha_k) }{ \sin ( 2 \alpha_k) } \frac{ \partial }{ \partial \alpha_k }.
\end{eqnarray}
As described above, we choose to ignore angular momentum in the ${\bm \eta}_k$ coordinates, so we can disregard the angular momentum operators $l_k^2$.  Moreover, only the leading term, which corresponds to two-body motion, of the recursion relation in Eq.~(\ref{eq:Lambda_k}) is relevant,
%DB I placed brackets around the aguments 2 alpha for easier readibility 
%\begin{align}
%\Pi^2 \equiv \Pi_{N-1}^2 = - \frac{ \partial ^2 }{ \partial \alpha^2 } 
%+ \frac{ (3N-9) - (3N-5) \cos 2 \alpha }{ \sin 2 \alpha } \frac{ \partial }{ \partial \alpha }.
%\label{eq:Pi2_definition}
%\end{align}
\begin{align}
\Pi^2 \equiv \Pi_{N-1}^2 = - \frac{ \partial ^2 }{ \partial \alpha^2 } 
+ \frac{ (3N-9) - (3N-5) \cos (2 \alpha) }{ \sin (2 \alpha) } \frac{ \partial }{ \partial \alpha }.
\label{eq:Pi2_definition}
\end{align}

The Schr\"odinger equation $H_{\rm{rel}} \Psi = E_{\rm{rel}} \Psi$, where $E_{\rm{rel}}$ is the energy of the relative motion, using Eq.~(\ref{eq:full_H_hyper}) represents a partial differential equation in $3(N-1)$ coordinates.  It is convenient at this point to introduce a set of basis functions defined on the hypersphere that diagonalize the fixed-$\rho$ Hamiltonian at each hyperradius $\rho$.  That is, the functions are eigenstates of hyperangular kinetic energy:
\begin{eqnarray}
\Lambda_{N-1}^2 Y_{\{ \lambda \}}(\rho; \Omega) = \epsilon_{\{ \lambda \}}(\rho) Y_{\{ \lambda \}}(\rho; \Omega).
\end{eqnarray}
The set $\{ \lambda \}$ represents a set of quantum numbers, here unspecified, that serve to distinguish the various basis states.  In the case of non-interacting particles, $a=0$, these are the usual hyperspherical harmonics and are independent of $\rho$ \cite{Smirnov77,Avery97_IJQC}.  In the present circumstance, however, eigenfunctions of $\Lambda_{N-1}^2$ are crafted subject to the Bethe-Peierls boundary conditions, in which case these functions
depend also parametrically on $\rho$, as we will see in the next section.      

We will refer to the functions  $Y_{\{ \lambda \}}$ as adiabatic basis functions.   Because they form a complete set on the hypersphere, it is possible to expand the 
%DB full 
relative
wave function as 
\begin{eqnarray}
\Psi = \rho^{-(3N-4)/2} \sum_{\{ \lambda \}} F_{\{ \lambda \}}(\rho) Y_{\{ \lambda \}}(\rho; \Omega)
\end{eqnarray}
for some set of radial expansion functions $F_{\{ \lambda \}}$.  Using this expansion in $H_{\rm{rel}} \Psi = E_{\rm{rel}} \Psi$ and projecting the resulting expression onto $Y_{\{ \lambda^{\prime} \}}$ yield a set of coupled equations, 
%DB I think I see why you used the lambda' here... when you project, you can continue working with lambda... however, the projection is not written out in the paper. So, I suggest to use lambda instead of lambda'...
%\begin{widetext}
%\begin{align}
%\sum_{\{ \lambda^{\prime} \}} \left[ - \frac{ \hbar^2 }{ 2m } \frac{ \partial^2 }{ \partial \rho^2 }
%+ \frac{ \hbar^2 }{ 2m } \left( \frac{ (3N-4)(3N-6) }{ 4 \rho^2 }
%+ \frac{ \Lambda_{N-1}^2 }{ \rho^2 } \right)
%+ \frac{ 1 }{ 2 } m \omega^2 \rho^2 - E_{rel} \right]
%F_{\{ \lambda^{\prime} \}}(\rho) Y_{\{ \lambda^{\prime} \}}(\rho; \Omega) = 0.
%\label{eq:Schrodinger_sum}
%\end{align}
%\end{widetext}
\begin{widetext}
\begin{align}
\sum_{\{ \lambda \}} \left[ - \frac{ \hbar^2 }{ 2m } \frac{ \partial^2 }{ \partial \rho^2 }
+ \frac{ \hbar^2 }{ 2m } \left( \frac{ (3N-4)(3N-6) }{ 4 \rho^2 }
+ \frac{ \Lambda_{N-1}^2 }{ \rho^2 } \right)
+ \frac{ 1 }{ 2 } m \omega^2 \rho^2 - E_{\rm{rel}} \right]
F_{\{ \lambda \}}(\rho) Y_{\{ \lambda \}}(\rho; \Omega) = 0.
\label{eq:Schrodinger_sum}
\end{align}
\end{widetext}
Here, the first term is a radial kinetic energy and the second term is an effective centrifugal energy that is a consequence of the hyperspherical coordinate system.

The set of coupled Eqs.~(\ref{eq:Schrodinger_sum}) is still exact, if all terms in the expansion are kept, but this is prohibitively expensive \cite{Avery_book}.  Instead, we follow common practice and make a Born-Oppenheimer-like approximation \cite{Blume00_JCP, Frey85_ChemPhys,Avery_book2,Starace79_PRA}. Namely, we assert that $\rho$ is a ``slow'' coordinate in the sense that we ignore the partial derivatives $\partial Y_{\{ \lambda \}}/\partial\rho$ in Eq.~(\ref{eq:Schrodinger_sum}).  When this is done, each term in the sum is independent of the others.  The solution representing the BEC is a single wave function specified by the quantum numbers $\{ \lambda \}$.  We adopt this approximation in what follows; the derivative couplings between adiabatic functions can be reinstated by familiar 
%DB means \cite{Blume00_JCP, Frey85_ChemPhys,Avery_book2,Starace79_PRA} 
means \cite{Blume00_JCP, Frey85_ChemPhys,Avery_book2,Starace79_PRA}. 
It is also worth noting that in the limit of infinite scattering length, the Born-Oppenheimer approximation again becomes exact \cite{Werner_PRA74}.

Here we will take this procedure one step further.  The particular adiabatic function of interest to us will describe two-body correlations, and will be chosen to reduce the collective set of quantum numbers $\{ \lambda \}$ to  a single quantum number $\nu$: 
\begin{eqnarray}
\label{eq:wavefxn}
\Psi = \rho^{-(3N-4)/2} F_\nu(\rho) Y_{\nu}(\rho;\Omega) 
%DB
.
\end{eqnarray}
The selection of this function will be described in the following section.  

This approximation is not unlike the  $K$-harmonic approximation, in which $Y$ is taken to be independent of all its arguments \cite{Bohn98_PRA}.  Such an approximation affords an easy calculation of ground state energies for small scattering lengths.   In this article we extend the definition of $Y_{\nu}$ to include a dependence on both hyperradius and on a restricted subset of hyperangles $\Omega$ that emphasize two-body correlations.  This additional flexibility will enable us to describe these correlations at any value of the scattering length. 

Using a single adiabatic function, the Schr\"odinger equation becomes a single ordinary differential equation in $\rho$: 
\begin{widetext}
\begin{align}
\left[ - \frac{ \hbar^2 }{ 2m } \frac{ d^2 }{ d \rho^2 } + V^{\rm diag}(\rho) \right] F_\nu(\rho)
+ \frac{ \hbar^2 }{ 2 m \rho^2 }  \langle \nu | \Lambda_{N-1}^2 | \nu \rangle
F_\nu(\rho) = E_{\rm{rel}} F_\nu(\rho),
\label{eq:adiabatic_equations}
\end{align}
\end{widetext}
where
\begin{eqnarray}
\label{eq:Vdiag}
V^{\rm diag}(\rho) = \frac{ \hbar^2 }{ 2m } \frac{ (3N-4)(3N-6) }{ 4 \rho^2 }
+ \frac{ 1 }{ 2 } m \omega^2 \rho^2
\end{eqnarray}
is the diagonal potential whose ground state 
%DB depicts  
supports
the non-interacting condensate wave function. The matrix element  $\langle \nu | \Lambda_{N-1}^2 | \nu \rangle$ representing an integral over the hypersphere is evaluated  in Appendix \ref{appendix:AngularKinetic}.  The additional term involving this matrix element  represents the additional kinetic energy in the many-body wave function due to the Bethe-Peierls boundary conditions.  It can be viewed as a kind of ``interaction'' potential, since the scattering length responsible for these boundary conditions arises ultimately from the two-body interaction.  

As a point of reference, the interaction term vanishes in the limit of zero scattering length. 
The noninteracting gas therefore has approximately the energy of the potential $V_{\rm diag}$ at its minimum, i.e., the energy of the gas is
\begin{align}
 V^{\rm diag}(\rho_0) &= \frac{\sqrt{(3N-4)(3N-6)}}{2} \hbar \omega, \\
 \label{eq:rho0}
 \rho_0 &= \left[ \frac{(3N-4)(3N-6)}{4} \right]^{1/4} a_{ho},
 \end{align}
where $ a_{ho}=\sqrt{\hbar/m \omega}$ is the characteristic trap length.

\section{Basis Functions}

In this section, we construct the adiabatic basis functions $Y_{\nu}$, focusing on the most relevant one to describe the gas-like state of the BEC.  We also construct the approximate matrix elements $\langle \nu | \Lambda_{N-1}^2 | \nu \rangle$ required to solve the hyperradial equation in~(\ref{eq:adiabatic_equations}).

\subsection{Jastrow Form and Pair Wave Function}

In describing 
%DB if not ``a'' then ``Bose gases''???
a 
dilute Bose gas with two-body interactions, for our present purposes, we are content with a simple description that includes only atom 
%DB pairs 
pair distances
$r_{ij}$, which explores only a tiny fraction of the available configuration space.  Specifically, consider a single pair described by $r_{12}$.  
%DB the following sentence seemed weird... at a minimum it was missing a bracket... here's an attempted rewrite...
%The relative motion of this pair (ignoring the angles, ${\hat r}_{12}$ is given by the single hyperangle $\alpha$ by $\sin \alpha = r_{12}/\sqrt{2} \rho$.  
Ignoring the direction of the pair distance vector $\mathbf{r}_{12}$, the relative motion of this pair is given by the single hyperangle $\alpha$ through $\sin \alpha = r_{12}/\sqrt{2} \rho$.
We can therefore contemplate a hyperangular basis function 
\begin{eqnarray}
\phi_{\nu}(\rho;\alpha) = \phi_{\nu}(\rho; \alpha_{12})
\end{eqnarray}
that is a function of only one of the $3N-4$ hyperangles, and that may depend parametrically on $\rho$.   

Of course, one may do the same for any 
%DB pair 
pair distance 
$r_{ij}$, and define a hyperangle via $\sin \alpha_{ij} = r_{ij}/\sqrt{2}\rho$.  Each such angle is expressed starting from a different set of Jacobi coordinates.  Starting from this nugget of a wave function, one can build a basis function that is appropriately symmetrized with respect to exchange of identical bosons, via 
\begin{eqnarray}
Y_\nu =  \frac{ \prod_{i<j} \phi_\nu(\rho; \alpha_{ij}) }{ \int d\Omega \sqrt{ \prod_{i<j} \phi_\nu(\rho; \alpha_{ij}) } }. 
\label{eq:Jastrow}
\end{eqnarray}
This form of the function as a product of all two-body contribution was made famous by Jastrow's pioneering effort \cite{Jastrow55_PR}.  It continues to find extensive use as a form of variational trial wave function, especially for Monte Carlo studies of many-body physics \cite{GreenBook}.  Note that at this point we deviate from the formalism of Ref.~\cite{Sorensen02_PRA}, which expresses symmetrization by means of a sum of two-body contributions (Faddeev approach) rather than a product (Jastrow approach).

The procedure for constructing and using adiabatic basis functions therefore consists of 1) choosing a reasonable set of functions $\phi_{\nu}$; and 2) living with the consequences of this choice.  To begin, the function $\phi_{\nu}$ should satisfy the Schr\"{o}dinger equation for the relative motion of two atoms when they are close to one another, that is, for small $\alpha$.  We define $\phi_{\nu}$ to be an eigenfunction of the differential operator 
%DB $\Pi^2$
$\Pi^2$,
%\begin{widetext}
\begin{align}
\Pi^2 \phi_{\nu} (\rho;\alpha) &= \epsilon_{\nu}(\rho) \phi_{\nu}(\rho;\alpha),
\label{eq:alpha_equation}
\end{align}
%\end{widetext}
where $\Pi^2$ is given in Eq.~(\ref{eq:Pi2_definition}).
The solution is determined from $\alpha=0$ (when the two particles coincide), up to a value of $\alpha = \alpha_d$ to be determined below.
To help visualize the consequences of this equation, it is sometimes useful to make the substitution 
\begin{eqnarray}
{\tilde \phi}_{\nu}(\rho;\alpha) = \sin \alpha \cos ^{(3N-7)/2} \alpha \phi_{\nu}(\rho;\alpha),
\end{eqnarray}
which satisfies the differential equation
\begin{align}
\left( - \frac{ \partial^2 }{ \partial\alpha^2 } - \frac{ 9N-19 }{ 2 } + \frac{ (3N-7)(3N-9) }{ 4 } \tan^2 \alpha \right)
{\tilde \phi}_{\nu} = \epsilon_{\nu} {\tilde \phi}_{\nu}. 
\end{align}
This version takes the form of an ordinary Schr\"odinger equation in $\alpha$, with a centrifugal potential energy term $\propto \tan^2 \alpha$ that confines the motion of the atom pair toward $\alpha=0$, and thus prevents the atom pair from getting too far apart.  This wave function therefore automatically emphasizes the action of this pair over the interaction of these atoms with others.  As the number of particles grows larger, the confinement is restricted to smaller values of $\alpha$.

The equation we will solve is, however, Eq.~(\ref{eq:alpha_equation}) with $\Pi^2$ given by Eq.~(\ref{eq:Pi2_definition}).  To solve it, we make the substitution $z = \sin^2 \alpha$, leading to 
\begin{align}
z(1-z) \frac{ d^2 \phi_{\nu} }{ dz^2 } + \left[ \frac{ 3 }{ 2 } 
+ \left( \frac{ 3 }{ 2 } - \frac{ 3N }{ 2 } \right) \frac{ d }{ dz } \right] \phi_{\nu} + \frac{ \epsilon_{\nu}}{ 4 } \phi_{\nu} = 0.
\end{align}
This equation has the form of the hypergeometric differential equation \cite{Erdelye53}, yielding two independent solutions for $\phi_{\nu}$, one regular and one irregular, in terms of the hypergeometric functions ${}_2F_1$:
\begin{align}
f_{\nu}(\alpha) &= {}_2F_1\left(-\nu, \frac{3N-5 }{ 2 } + \nu, \frac{ 3 }{ 2 }; 
\sin^2 \alpha \right) \nonumber \\
g_{\nu}(\alpha) &= (\sin \alpha)^{-1} {}_2F_1 \left( - \nu - \frac{ 1 }{ 2 }, \frac{ 3N-6}{ 2 } + \nu,
\frac{ 1 }{ 2 }; \sin^2 \alpha \right).
\end{align}
Here, $\nu$ is a to-be-determined index that will in turn determine the eigenvalue,
\begin{align}
\Pi^2 \left\{ \begin{array}{c} f_{\nu} \\ g_{\nu} \end{array} \right\}
 = \epsilon_{\nu}  \left\{ \begin{array}{c} f_{\nu} \\ g_{\nu} \end{array} \right\}
= 2\nu(2 \nu + 3N-5) \left\{ \begin{array}{c} f_{\nu} \\ g_{\nu} \end{array} \right\}.
\label{eq:evalue}
\end{align}
A perfectly general solution to Eq.~(\ref{eq:alpha_equation}) is then
\begin{eqnarray}
A f_{\nu} + B g_{\nu}.
\end{eqnarray}
In general, both constants $A$ and $B$, as well as the index $\nu$, will depend on the hyperradius and the scattering length, as we will now show.

\subsection{Boundary Conditions}

The coefficients $A$ and $B$ are determined by applying boundary conditions.  The first such condition  occurs at $\alpha=0$, where the two atoms meet, and is given by the Bethe-Peierls condition (\ref{eq:BP_boundary_conditions}).  We first write this condition in hyperspherical coordinates. We have, in the limit of small $\alpha$, and for fixed hyperradius $\rho$,
\begin{align}
\frac{ 1 }{ r_{12} \phi_{\nu} } \frac{ \partial (r_{12} \phi_{\nu} ) }{ \partial r_{12} } &=
\frac{ 1 }{ \sqrt{2} \rho \sin \alpha \phi_{\nu} } 
\frac{ \partial ( \sqrt{2} \rho \sin \alpha \phi_{\nu} )}{ \partial (\sqrt{2} \rho \sin \alpha) } \nonumber \\
& \approx  \frac{ 1 }{ \sqrt{2} \rho (\alpha \phi_{\nu}) } \frac{ \partial (\alpha \phi_{\nu}) }{ \partial \alpha } 
= - \frac{ 1 }{ a }.
\end{align}
Next, expanding the hypergeometric functions near $\alpha \approx 0$ gives $\phi_{\nu} \approx A   + B/\alpha$, whereby
\begin{align}
- \frac{ \sqrt{2} \rho }{ a } = 
\frac{ 1 }{ \alpha (A + B/ \alpha) } \left.\frac{ \partial [ \alpha (A+B/ \alpha)]  }{ \partial \alpha } \right\vert_{\alpha=0} = 
\frac{ A }{ B },
\label{eq:BC_alpha0}
\end{align}
which determines the ratio of the coefficients. Note that this ratio depends on $\rho$.   It is significant that this boundary condition is applied to a single pair of 
%DB the construction seemed awkward...
%DB particles, then is 
particles and is then
implicitly applied to all pairs by the form of the wave function in Eq.~(\ref{eq:Jastrow}).  This is in accord with the notion that the Bethe-Peierls boundary condition is  local, and influences each pair independently of what the other pairs are doing.  

The other boundary condition on $\phi_{\nu}$ is inspired by the brilliant reinterpretation of the Jastrow wave functions by Pandharipande and Bethe \cite{Pandharipande73_PRC,Pandharipande77_PRA}.  In this version $\phi_{\nu}$ is viewed as a piece of the pair-correlation function, related to the probability of finding this pair 
a given distance apart.  To this end, $\phi_\nu$ is required to  approach unity on an appropriate length scale $r_d$, which in hyperspherical 
%DB coordinate 
coordinates
we translate into an appropriate hyperangular scale $\alpha_d$.  Beyond this characteristic scale, the atoms are assumed to be uncorrelated
%DB , so 
and
the wave function is required to satisfy
\begin{align}
\phi_{\nu}( \rho; \alpha) &= 1, \;\;\;\; \alpha > \alpha_d, \nonumber \\
\left. \frac{ \partial \phi_{\nu} }{ \partial \alpha } \right\vert_{\alpha_d} &= 0.
\label{eq:BC_alphad}
\end{align}
Setting the function to unity for $\alpha > \alpha_d$ is convenient but arbitrary.  

There remains the issue of determining a reasonable value of $\alpha_d$.  This is also done by viewing $|\phi_{\nu}|^2$ as a pair-correlation function.  Given the location of atom 1, $\alpha$ encodes the distance to the next atom, 2.  If $\alpha$ becomes too large, then atom 2 can go explore parts of the hypersphere where a third atom is likely to be found.  At that point, the description in terms of a pair-correlation function is not so useful, and $\phi_{\nu}$ should not be extended non-trivially this far.

Therefore, $\alpha$ should be limited to a region of the hypersphere where, on average, only one atom will be found in addition to the fixed atom 1.  This is the essence of  the lowest-order constrained variational, or  LOCV, approximation \cite{Pandharipande73_PRC}.  Given the symmetrized wave function $Y_\nu$ in Eq.~(\ref{eq:Jastrow}),  the average number of atoms within $0 \le\alpha \le \alpha_d$ of the atom presumed to lie at $\alpha=0$ is 
\begin{align}
(N-1) \frac{  4 \pi \int_{0}^{\alpha_d} d\Omega_{\alpha} \int d\Omega_{N-2}
\prod_{i<j} |\phi_\nu(\rho;\alpha_{ij})|^2 }
{ \int d\Omega_{N-1}
\prod_{i<j} |\phi_\nu(\rho;\alpha_{ij})|^2 },
\label{eq:LOCV_integral}
\end{align}
which returns $N-1$ when $\alpha_d = \pi/2$.  The expression given in Eq.~(\ref{eq:LOCV_integral}) can be written as 
\begin{align}
(N-1) \frac{ 4 \pi \int_0^{\alpha_d} d \Omega_{\alpha} h_\nu(\alpha) }
{ 4 \pi \int_0^{\pi/2}  d \Omega_{\alpha} },
\label{eq:average_number}
\end{align}
where $h_{\nu}$ is defined through
\begin{align}
h_{\nu}(\alpha) = \left( 4 \pi \int d\Omega_{\alpha} \right)
\frac{ \int d\Omega_{N-2} \prod_{i<j} |\phi_{\nu}(\rho;\alpha_{ij})|^2 }
{ \int d\Omega_{N-1} \prod_{i<j} |\phi_{\nu}(\rho;\alpha_{ij})|^2 }.
\end{align}
This is a difficult multidimensional integral to evaluate, but it is often conveniently expanded into powers of integrals of the (presumed small) quantities $1 - |\phi_{\nu}(\rho;\alpha_{ij})|^2$.  The lowest-order term of this expansion, and the approximation we will use here, then gives the approximation
\begin{align}
h_{\nu}(\alpha) = |\phi_{\nu}(\rho;\alpha)|^2.
\label{eq:correlation_approximation}
\end{align}
Using this approximation, and setting the average number of atoms (\ref{eq:average_number}) to unity, yields the normalization criterion
\begin{align}
&\int_0^{\alpha_d} d \alpha \sin^2 \alpha \cos^{3N-7} \alpha |\phi_{\nu}(\rho;\alpha)|^2 \nonumber\\
&\qquad \qquad= \frac{ 1 }{ N-1 } \int_0^{\pi/2} d \alpha \sin^2 \alpha \cos^{3N-7} \alpha \nonumber \\
&\qquad \qquad= \frac{ 1}{ N-1 } \frac{ \sqrt{\pi} }{ 4 } \frac{ \Gamma( 3N/2-3) }{ \Gamma( 3N/2 - 3/2 ) }. 
\label{eq:BC_norm}
\end{align}
This requirement must be met self-consistently.  That is, the boundary conditions (\ref{eq:BC_alpha0}) and (\ref{eq:BC_alphad}) determine $\phi_{\nu}$ for any given $\alpha_d$; but $\alpha_d$ must also be chosen so that (\ref{eq:BC_norm}) is satisfied.  Notice that the right-hand-side of (\ref{eq:BC_norm}) scales as $N^{-5/2}$ as $N$ gets large, whereby $\alpha_d$ gets smaller with increasing $N$.  Any pair of atoms must be closer together to avoid the other atoms, when there are more atoms.

The LOCV approach has been employed successfully  when Jastrow wave functions are used.  Usually, one posits an unknown two-body correlation function, to be determined variationally, minimizing some energy while varying the parameters of the trial function.  This method has also been employed, in Cartesian coordinates, to describe the energetics of a homogeous Bose gas with large scattering length \cite{Cowell02_PRL,Rossi_PRA89,Kalas_PRA}.  Used in the present context of hyperspherical coordinates, we dub this approach the hyperspherical LOCV (H-LOCV) method.  

Putting together the boundary conditions, we have
\begin{eqnarray}
A - \frac{ \sqrt{2} \rho }{ a } B &=& 0 \nonumber \\
f_{\nu}^{\prime}(\alpha_d)A +  g_{\nu}^{\prime}(\alpha_d)B &=& 0,
\end{eqnarray}
where  $f_{\nu}^{\prime}(\alpha_d) \equiv (df_{\nu}/d \alpha )|_{\alpha_d}$, and similarly for $g_{\nu}^{\prime}(\alpha_d)$, can be determined from the derivatives of the hypergeometric functions.  This system of equations for $A$ and $B$ can be solved only if
\begin{eqnarray}
\det \left\vert \begin{array}{cc} 1 & -\sqrt{2} \rho / a \\  f_{\nu}^{\prime}(\alpha_d) & g_{\nu}^{\prime}(\alpha_d)
\end{array} \right\vert = 0.
\label{eq:determinant}
\end{eqnarray}
Therefore, given $a$ (which defines the physical problem to be solved) and $\rho$ (which defines the hyperradius at which the adiabatic function 
%DB $\phi_{\nu}$ 
$Y_{\nu}$
is desired), zeroing the determinant (\ref{eq:determinant}) determines a value of $\nu$ for any given $\alpha_d$, which is then varied to self-consistently satisfy the normalization.

\begin{figure}[htbp!]
\centering
\begin{subfigure}[]{}
\centering
\includegraphics[width=.48\textwidth]{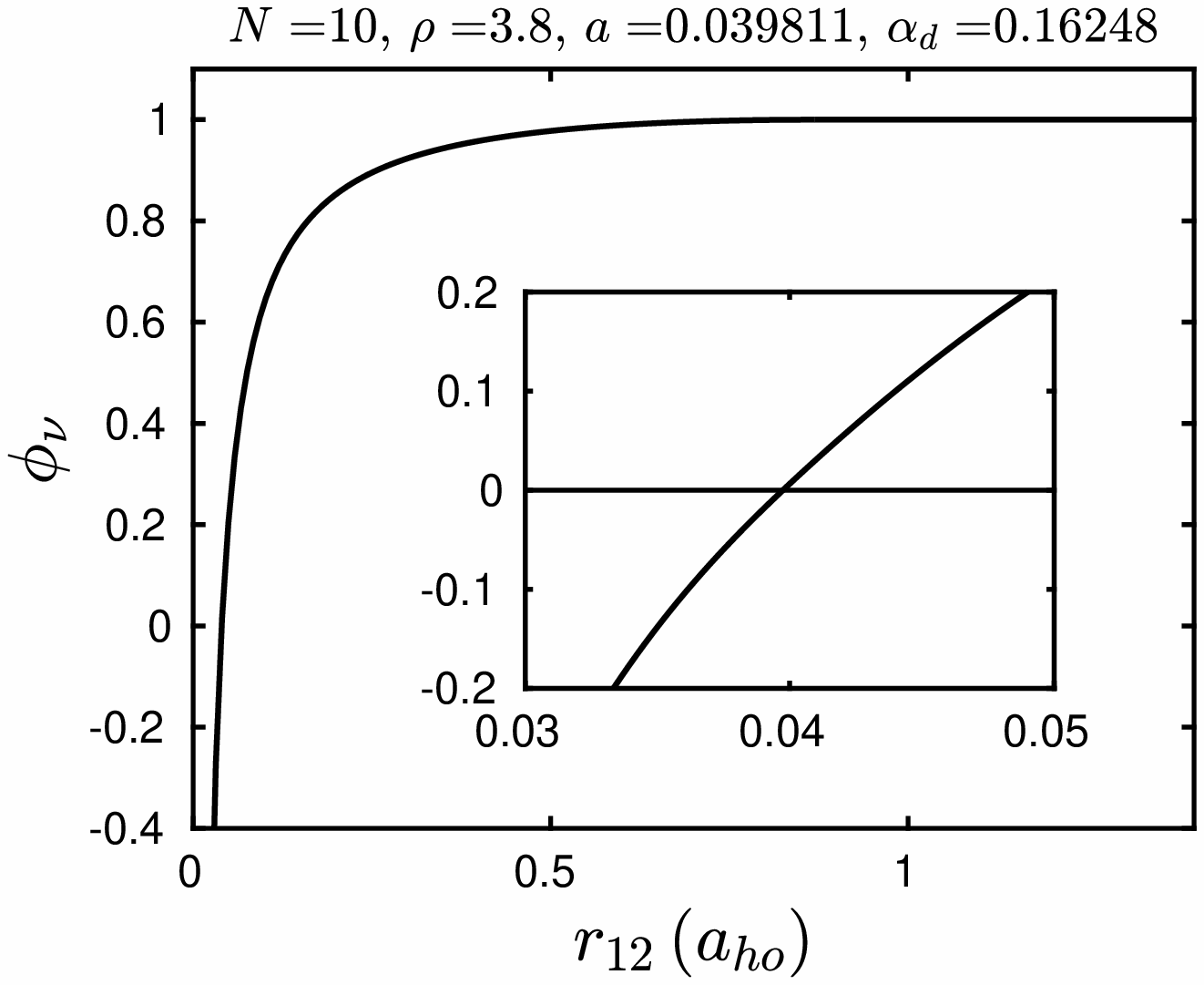}
%\caption{\label{fig:Wavefxn_N=10smallasc}$a=0.0398$, $\rho=3.8$, $\alpha_d=0.1625$}
\end{subfigure}\quad
\begin{subfigure}[]{}
\centering
\includegraphics[width=.48\textwidth]{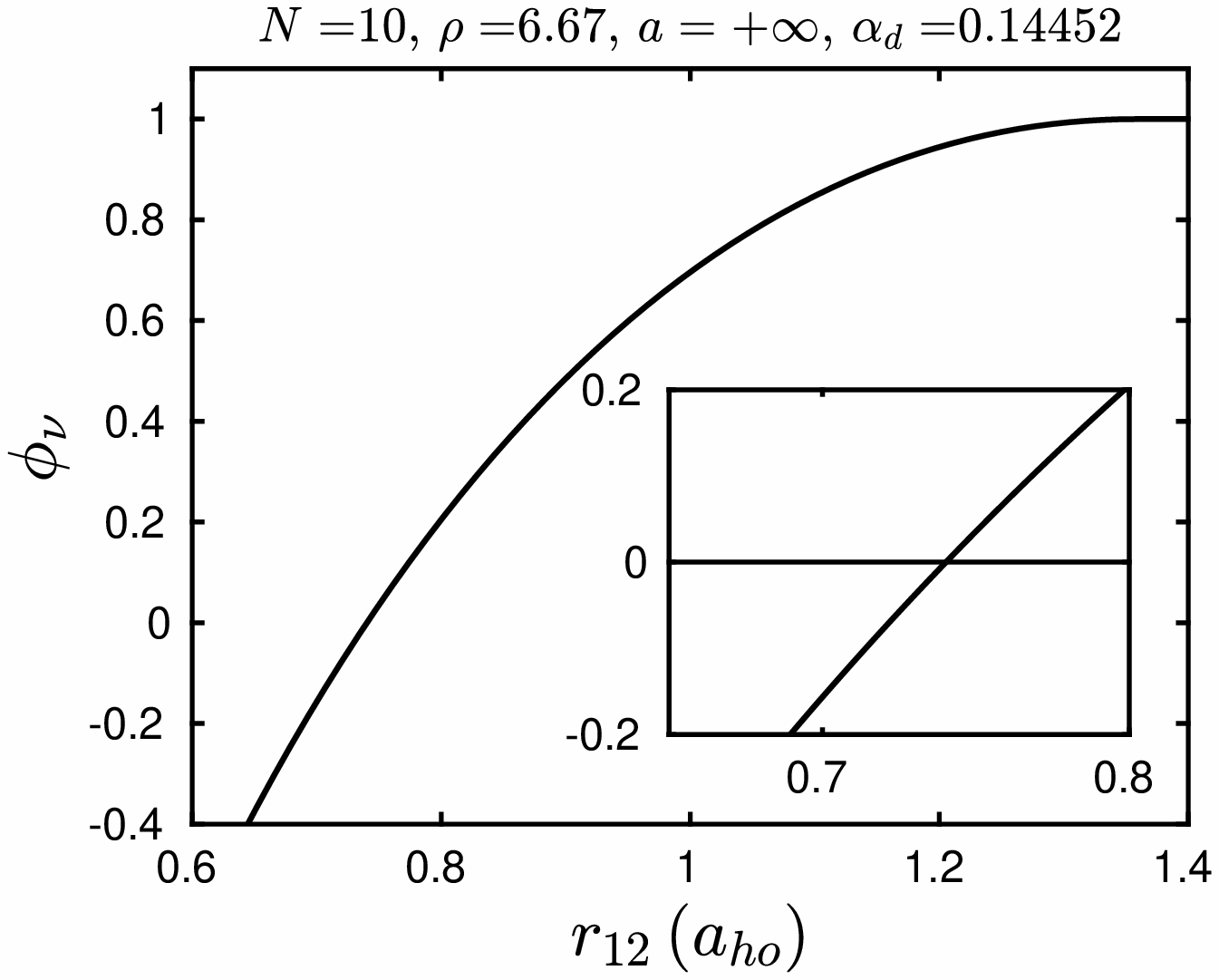}
%\caption{\label{fig:Wavefxn_N=10largeasc}$a=0.0398$, $\rho=3.8$, $\alpha_d=0.1625$}
\end{subfigure}
\caption{\label{fig:Wavefxn_N=10} The angular wave function 
%DB $\phi$ 
$\phi_{\nu}$
as a function of $r_{12}$ for $N=10$ at (a) small scattering length and on (b) resonance $a=\infty$. All length scales are in units of $a_{ho}=\sqrt{\hbar/m\omega}$. The insets show the zero-crossings of the curves. The crossing occurs at $\approx a$ for small scattering length and at some finite value in the 
%DB unitarity 
unitary
limit. Note that 
%DB $\phi$ 
$\phi_{\nu}$
is actually a function of the hyperangle $\alpha$. The relation $r_{12}=\sqrt{2} \rho \sin{\alpha}$ is used to convert the angle $\alpha$ to the 
%DB pair 
pair distance
$r_{12}$.  Here $\rho$ has its value at the minimum of the corresponding hyperradial potential.  }
\end{figure}

%DB \subsection{Renormalized scattering length}
\subsection{Renormalized Scattering Length}

The procedure outlined above generates an entire spectrum of $\nu$ values denoting various pair excitations of the condensate.  The lowest member of this spectrum, with no nodes, represents a self-bound liquid-like state \cite{Gao05_PRL,Kalas_PRA}, and is not what we are interested in here.  Rather, for positive scattering length, the BEC wave function in the relative coordinate $r_{12}$ should contain a single node to describe the gas-like BEC ground state.  

We denote the location of this node by $a_c$ (the subscript ``c'' denoting the value of $r_{12}$ where the wave function crosses zero).  For small scattering length, this node lies at a distance $a_c \approx a$ as demonstrated in Fig.~\ref{fig:Wavefxn_N=10}(a).   However, as $a$ gets larger this node, confined to a hyperangular range $0 \le \alpha \le \pi/2$, must saturate, leading to  a finite $a_c$ for any finite hyperradius. The saturation value of $a_c$, termed $a^*$, represents the effective scattering length on resonance.    This saturation is illustrated in Fig.~\ref{fig:Wavefxn_N=10}(b), showing $\phi_{\nu}(\alpha)$ for $a=\infty$. 

\begin{figure}[!htbp]
\centering
\includegraphics[scale=0.6]{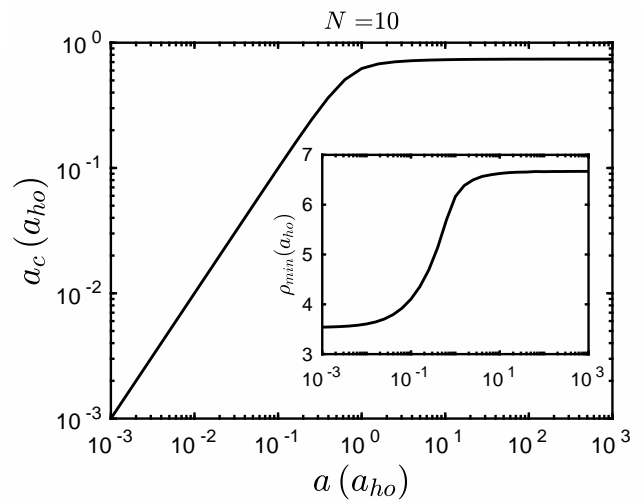}
\caption{\label{fig:ZeroCrossing_N=10} The zero-crossing $a_c$ of the 
%DB hyperangular 
angular
wave function 
%DB $\phi$ 
$\phi_{\nu}$
for $N=10$ as a function of the scattering length $a$. The inset shows the value $\rho_{min}$ that is used to compute $a_c$; $\rho_{min}$ is taken to be the value of $\rho$ where the effective potential 
%DB $V_{eff}$, 
$V^{\rm eff}$, 
discussed in the next section, is at its minimum. }
\end{figure}

 Figure~\ref{fig:ZeroCrossing_N=10} tracks the value of the length $a_c$ over the entire range of positive scattering 
%DB length.  
lengths.
For small $a$,  $a_c$ grows linearly.   It then rolls over, on length scales comparable to the harmonic oscillator length, to saturate to a value $a^*$ when $a=\infty$.  This auto-renormalization of the scattering length is inherent in the H-LOCV method and independent of any local density approximation of the gas. 

To summarize: the adiabatic eigenfunction $\phi_{\nu}(\rho; \alpha)$ that we seek has the following properties: 1) it satisfies the differential equation (\ref{eq:alpha_equation}) in $0\le \alpha \le \alpha_d$; 2) it satisfies boundary conditions (\ref{eq:BC_alpha0}) and (\ref{eq:BC_alphad}) at a suitable $\alpha_d$, chosen so that 3) the normalization (\ref{eq:BC_norm}) is satisfied; and 4) the wave function that results has a single node in $\alpha$.  The algorithm to find such a function is not terribly complicated, inasmuch as the wave function can be written analytically in terms of hypergeometric functions.  This procedure yields a single wave function with a particular value of the index $\nu$, for each value of scattering length and hyperradius. For a single channel potential, the effective potential $V^{\rm eff}(\rho)$ corresponding to this $\nu$ is given by (see Appendix A)
\begin{align}
\label{eq:Veff}
V^{\rm eff}(\rho) &= V^{\rm diag}(\rho) +  V^{\rm int}(\rho),
\end{align}
where $V^{\rm diag}(\rho)$ is given in Eq.~(\ref{eq:Vdiag}), and the ``interaction'' potential, derived in Appendix A, is given by
\begin{equation}
 V^{\rm int}=\frac{ \hbar^2 }{ 2m \rho^2}\frac{ N }{ 2 } 2\nu(2\nu+3N-5).
\end{equation}

\begin{figure}[!htbp]
\centering
\includegraphics[scale=0.6]{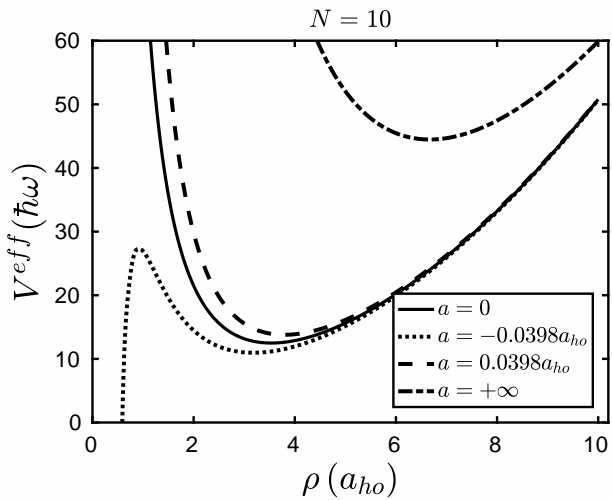}
\caption{\label{fig:VeffVsRho} The effective potential 
%DB $V_{eff}$ 
$V^{\rm eff}$ 
as a function of the hyperradius $\rho$ for $a=0$ (noninteracting case), $a>0$ (repulsive interaction), and $a<0$ (attractive interaction). For any $a>0$ and $N$, a local minimum always exists. However, for a given $a<0$, there exists a maximum $N$ when the local minimum of 
%DB $V_{eff}$ 
$V^{\rm eff}$ 
starts to disappear. The highest 
%DB $V_{eff}$ 
$V^{\rm eff}$ 
curve corresponds to 
%DB
the 
$a\rightarrow +\infty$ case.}
\end{figure}

The effect of this interaction potential can be seen in Fig.~\ref{fig:VeffVsRho}.  
When $a=0$ (solid curve), the $\rho^{-2}$ kinetic behavior dominates for small $\rho$ while $V^{\rm eff}$ takes on the $\propto \rho^2$ behavior of the trapping potential as $\rho \rightarrow \infty$. The ground state solution to the noninteracting case is just the well known solution to the quantum harmonic oscillator problem with 
%DB
relative 
energy 
%DB $3(N-1)/2\hbar \omega$. 
$3(N-1)\hbar \omega/2$. 
For small positive $a$ (dashed line), $V^{\rm eff}$ rises above the noninteracting $V^{\rm eff}(a=0)$, and the local minimum $\rho_0(a)$, where the condensate is centered, increases, indicating an expansion in the overall size of the condensate. This behavior is consistent with the repulsive nature of the contact interaction characterized by a positive scattering length. For small negative scattering length (dotted line), the opposite is true: the atoms pull in towards each other due to the attractive contact interaction; hence the decrease in energy and the condensate size.  

Such features of $V^{\rm eff}$  have been illustrated previously using the K-harmonic method \cite{Bohn98_PRA}. In the K-harmonic method, however, $V^{\rm int}$ is proportional to the scattering length, whereby this method suffers the same limitation to small-$a$ as does mean-field theory.  In the hyperspherical LOCV method, by contrast, the effective scattering length saturates and the effective potential remains finite even in the $a \rightarrow \infty$ limit.  This $V^{\rm eff}$ is the dash-dotted curve in Fig.~\ref{fig:VeffVsRho}.

%For negative scattering length, a trapped BEC is known to be stable for small $|a|$ and small $N$. In the hyperspherical picture,  when $a<0$,  the attractive $\rho^{-3}$ contribution from $V^{\rm in}$ weakens the repulsive barrier at small $rho$.  There can only be a metastable condensate state until the gas tunnels through the barrier to the region of small $\rho$ where three-body recombination prevails and destroys the gas. In fact, this metastable state can only exist for small values of negative $a$ as the system becomes large. That is, it takes only a small amount of $|a|$ to break down the centrifugal barrier as $N$ grows. Figure~\ref{fig:NcVsScatteringLength} shows the critical number, $N_C$, of bosons that can be condensed for a given negative scattering length. The solid line is the analytic prediction produced by the K-harmonic method \cite{Bohn98_PRA} and  a variational method using Gaussian functions \cite{Dalfovo99_RevModPhys}, while the discrete data points come from our LOCV computation. Our numerical results are in excellent agreement with the analytic approximation. 

% \langle \nu | \Lambda_{N-1}^2 | \nu^{\prime} \rangle
% = \frac{ N }{ 2 } 2\nu(2\nu+3N-5) \delta_{\nu \nu^{\prime}}

%%%%%%%%%%%%%%%%%%%%%%%%%%%%%%%%%%%%%

\section{Ground State Properties in the Hyperspherical-LOCV Approximation}

In this section, we report on ground state properties of the BEC in the H-LOCV approximation.  Recall that the method begins by separating the center of mass energy $(3/2)\hbar \omega$, then solves for the relative energy $E_{\rm rel}$ using the Hamiltonian (\ref{eq:full_H_hyper}).  In the results of this section, we report the full condensate ground state 
%DB
energy 
$E_0 = E_{\rm rel} + (3/2)\hbar \omega$.  

%DB \subsection{General features of the ground state}
\subsection{General Features of the Ground State}

\begin{figure}[!htbp]
\centering
\includegraphics[scale=0.6]{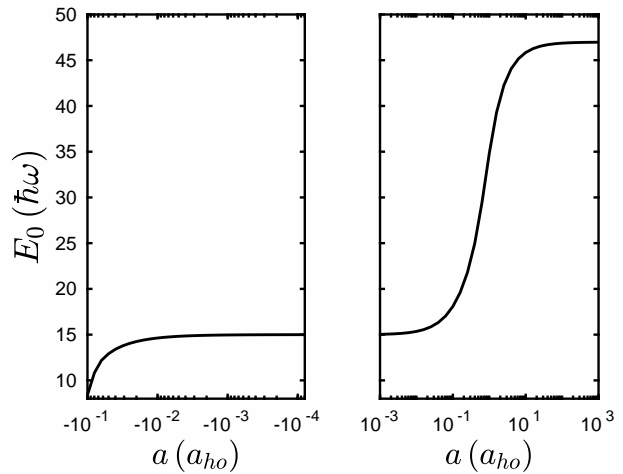}
\caption{\label{fig:GroundStateEnergyVsScatteringLength_N=10} Ground state energy $E_0$ as a function of the scattering length $a$ for $N=10$. The left and right panels consider negative and positive scattering lengths.}
\end{figure}

 The energy per particle versus scattering length is shown  in Fig.~\ref{fig:GroundStateEnergyVsScatteringLength_N=10} for $N=10$.  On the left is the energy for attractive interaction $a<0$, on the right for repulsive interaction $a>0$.    The $E_0(a<0)$ curve connects smoothly with $E_0(a>0)$ at $a=0$ with 
%DB $E_0=3N/2\hbar \omega$, 
$E_0=3N\hbar \omega/2$, 
then increases smoothly until it saturates in the large-$a$  limit. A similar behavior can be observed for any $N$.

As is well known, a trapped gas is mechanically stable only for negative scattering length of small magnitude.  In a hyperspherical picture such as this one, a collapse instability occurs when the attractive $1/\rho^3$ interaction potential overcomes the repulsive $1/\rho^2$ centrifugal potential and the effective potential $V^{\rm eff}$ in Fig.~\ref{fig:VeffVsRho} has no classical inner turning point.  On the left of Fig.~\ref{fig:GroundStateEnergyVsScatteringLength_N=10}, the collapse region occurs near $a\approx-0.1\, a_{ho}$ where a metastable condensate would cease to exist. 

Generally, this means that for a harmonically trapped Bose condensate, for any negative value of $a$ only a certain critical number $N_c$ of atoms can be contained before collapse occurs. This is an intrinsically small-$|a|$ phenomenon, and was quantified in hyperspherical terms in Ref. \cite{Bohn98_PRA}.  Our H-LOCV results are in agreement with this calculation, finding that 
%DB $N_c \sim 0.671/|a|$.   
$N_c \sim 0.671\, a_{ho}/|a|$.   

%DB \subsection{Positive scattering length}
\subsection{Positive Scattering Length}

\begin{figure}[!htbp]
\centering
\includegraphics[scale=0.65]{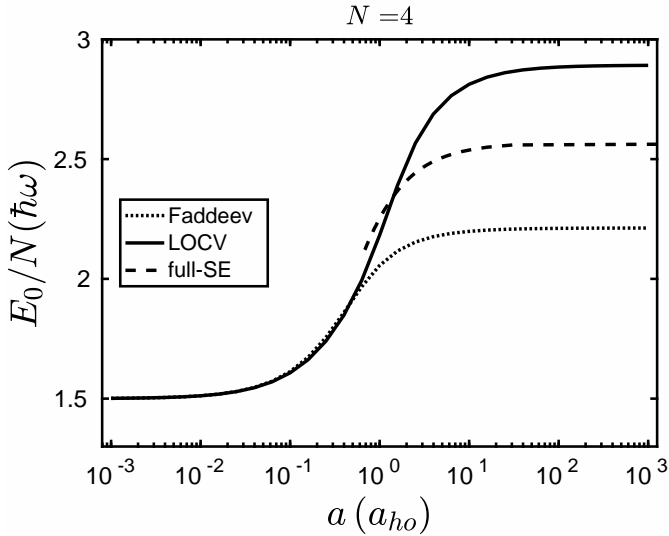}
\caption{\label{fig:GroundStateEnergy_N=4} Ground state energy per particle $E_0/N$ as a function of the scattering length $a$ for $N=4$. }
\end{figure}

\begin{figure}[!htbp]
\centering
\includegraphics[scale=0.65]{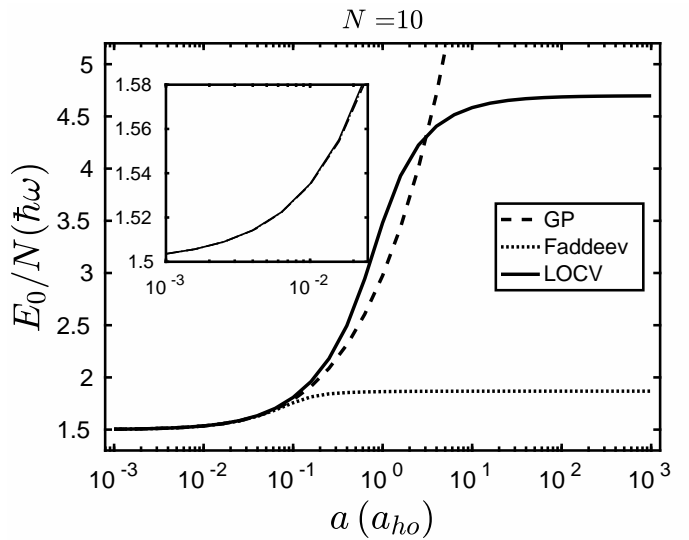}
\caption{\label{fig:GroundStateEnergy_N=10} Ground state energy per particle $E_0/N$ as a function of the scattering length $a$ for $N=10$.
%DB
The inset shows a blow-up of the weakly interacting regime.}
\end{figure}

For the rest of this paper we will focus on the positive scattering length case.  To this end, the ground state energy of the condensate is shown versus scattering length, for gases of $N=4$ and $N=10$ particles, in Figs.~\ref{fig:GroundStateEnergy_N=4} and~\ref{fig:GroundStateEnergy_N=10}.   The H-LOCV result is shown as a solid line.

In Fig.~\ref{fig:GroundStateEnergy_N=4}, for $N=4$ atoms, the energy per particle in the H-LOCV method  saturates at a finite value in the resonant limit, just over  $2.5\,\hbar \omega$ per particle.  This case, with $N=4$, can be compared directly to an accurate numerical solution to the full four-particle Schr\"odinger equation, and is therefore an important case to check.

The 
%DB direct 
full
numerical calculation incorporates a two-body potential of the form $v_0 \exp[ - (r_{jk}/\sqrt{2}r_0)^2]$, with range $r_0 = 0.025\, a_{ho}$ and depth $v_0$ adjusted to achieve the desired two-body scattering length.  This model also incorporates a repulsive three-body Gaussian potential to eliminate deep-lying bound states of the system.  Within this model, an energy spectrum is calculated using a basis set of correlated Gaussian functions.  Further details are provided in \cite{Blume_preprint}.  

The numerical spectrum includes a great many states that represent bound cluster states, the analogues of Efimov states for the trapped system.  They are characterized by, among other things, a dependence on the three-body potential.
%DB
%By contrast, the nearly-universal state corresponding to the gas-like BEC groun%d state is identified by its near independence from the three-body potential, a%s well as by I'M NOT SURE WHAT IS MEANT BY THE
%FOLLOWING??? projecting onto nonzero hyperradius.
By contrast, the nearly-universal state corresponding to the gas-like BEC
ground state is identified by its near independence from the
three-body potential and its vanishingly small amplitude
at small hyperradii.
The energy of the corresponding state is shown as a dashed line in Fig.~\ref{fig:GroundStateEnergy_N=4}.  The comparison shows that the H-LOCV method gets the saturation value of the energy per particle approximately right, at least for $N=4$ particles.  

Also shown in Fig.~\ref{fig:GroundStateEnergy_N=4} (dotted line) is the result of an alternative hyperspherical method that symmetrizes the wave function using a {\it sum} of two-body terms (Faddeev method) rather than a product as we use here in the H-LOCV method.  The Faddeev method is accurate for three atoms in a trap \cite{Jonsell02_PRL,Braaten06_PhysRep},  and the comparison in Fig. \ref{fig:GroundStateEnergy_N=4} suggests that it is viable for four particles as well.

A difference occurs, however, in Fig.~\ref{fig:GroundStateEnergy_N=10}
%DB , 
for $N=10$ atoms.  Here a full numerical solution to the Schr\"odinger equation is not available, but we can compare the H-LOCV and Faddeev hyperspherical methods side by side.  The solid line represents the results from the H-LOCV approximation.  Also shown are the results from the mean-field Gross-Pitaevskii (GP, dashed line) \cite{Pethick} and hyperspherical-Faddeev (Faddeev, dotted line) models \cite{Sogo05_JPB,Sorensen02_PRA,Sorensen03_PRA,Sorensen04_JPB}. If we zoom in to the $a/a_{ho}\ll1$ domain (inset), we find that all models agree well in the weakly interacting regime. In addition to these, other independent computations such as the K-harmonic  method \cite{Bohn98_PRA} and diffusion Monte-Carlo \cite{Blume01_PRA} have predicted the weakly interacting system accurately. 

In the large scattering length limit, Fig.~\ref{fig:GroundStateEnergy_N=10} shows the saturation of the energy per particle to an asymptotic value, which grows with $N$ (as we will show below, the energy per particle scales as $N^{1/3}$ as anticipated by 
%DB
the 
Thomas-Fermi approximation \cite{Ding17_PRA}). It is well-known that the GP approximation diverges in this limit.  In addition, Monte Carlo calculations with hard spheres cannot approach this limit, although Monte Carlo calculations with renormalized scattering lengths can circumvent this issue \cite{Rossi_PRA89}.   

Of particular interest is  the hyperspherical Faddeev method (dotted line).  Using the formalism of Ref. \cite{Sogo05_JPB}, one finds that the energy per particle saturates to  a far lower value than does the H-LOCV. This situation gets worse as the number of particles $N$ increases.  Note that the H-LOCV $\nu$ that goes into the $V^{\rm int}$ plays a different role from the pure hyperspherical-Faddeev $\nu$ \cite{Sogo05_JPB,Sorensen02_PRA}; also, the latter does not place any restriction on the $\alpha$ domain. Using the formalism of Ref. \cite{Sogo05_JPB,Sorensen02_PRA}, we  find that the asymptotic hyperspherical-Faddeev $\nu$ at large $N$ approaches $\nu\rightarrow2$ in the large-$a$ limit.
In this model, therefore, the energy per particle is a decreasing function of $N$, and does not adequately describe the system in this limit.  

\begin{figure}[htbp!]
\centering
\begin{subfigure}[]{}
\centering
\includegraphics[width=.5\textwidth]{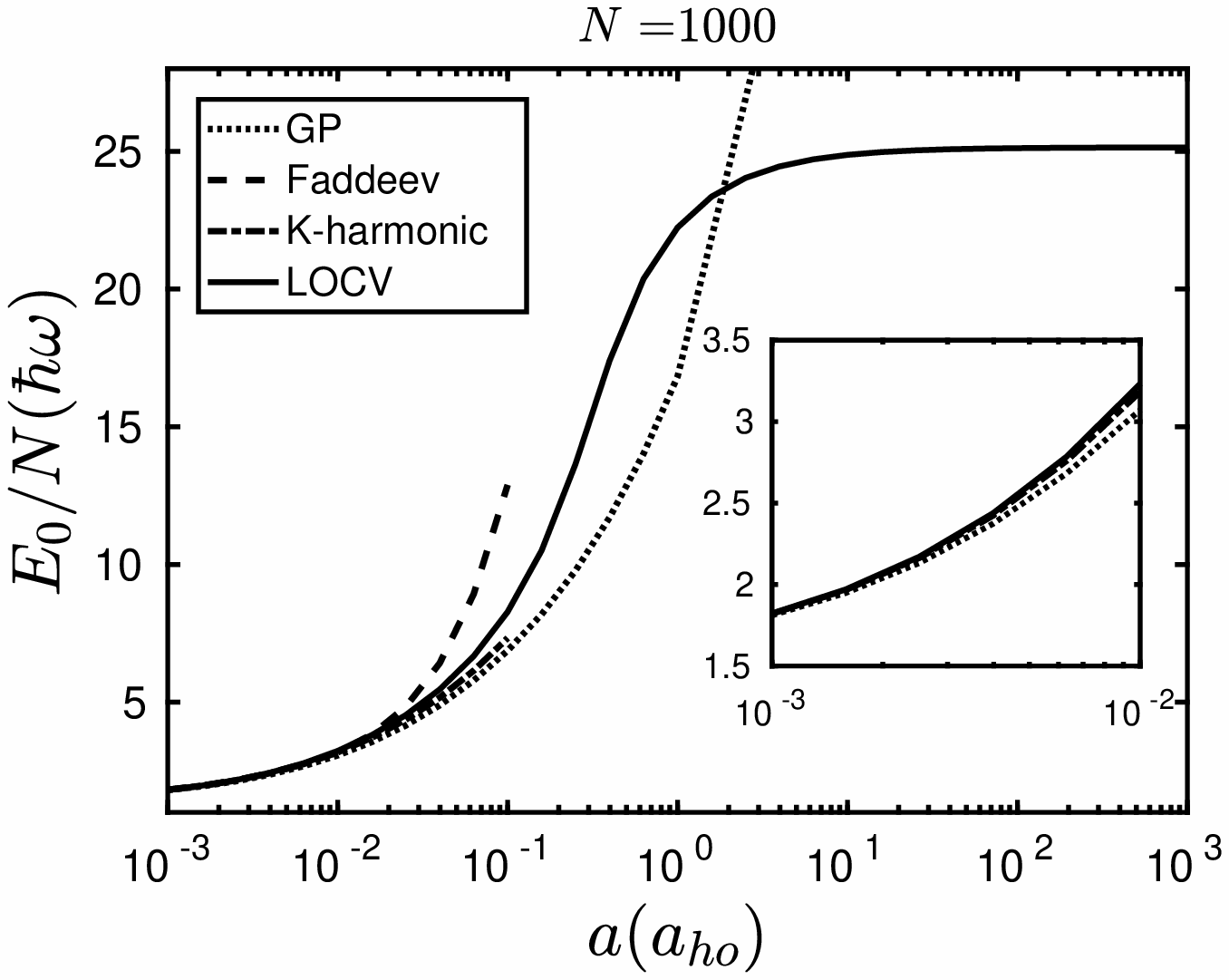}
%\caption{$a=0.0398$, $\rho=3.8$, $\alpha_d=0.1625$}
\end{subfigure}\quad
\begin{subfigure}[]{}
\centering
\includegraphics[width=.5\textwidth]{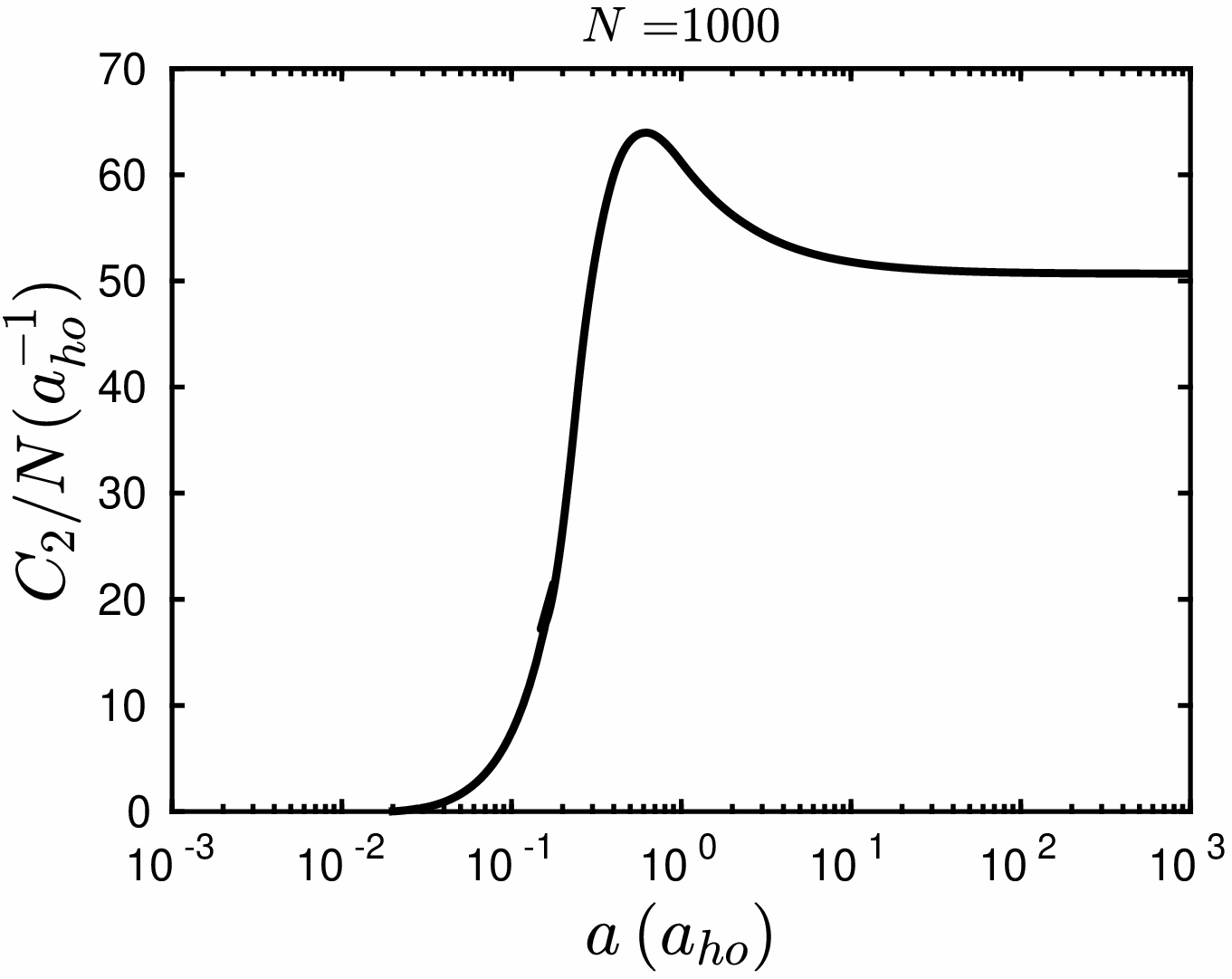}
%\caption{$a=0.0398$, $\rho=3.8$, $\alpha_d=0.1625$}
\end{subfigure}
\caption{
%DB (a) Ground state energy per particle $E_0/N$ as a function of the scattering length $a$ for $N=1000$. 
%DB (b) ground 
(a) Ground state energy per particle, in units of $\hbar \omega$. The inset shows the various perturbative versions (small $a$ region). %DB the 
(b) Two-body contact per 
%DB particle , 
particle,
in units of $a_{ho}^{-1}$, as a function of 
%DB
the
scattering length 
%DB
$a$
for $N=1000$.}
\label{fig:Contact_N=1000}
\end{figure}

One distinct benefit of the H-LOCV approach is the ease with which it extends to the large-$N$ limit.  In this formalism, $N$ is just a parameter in the differential equations and their boundary conditions.  For example,  Fig.~\ref{fig:Contact_N=1000}(a) shows the energy per particle for $N=1000$ particles. Obtaining these results is not computationally harder than the same result for $N=10$.  Also shown, in Fig.~\ref{fig:Contact_N=1000}(b), is the two-body contact,  $C_2$, a thermodynamic quantity given by \cite{Smith14_PRL}
\begin{equation}
 C_2=8\pi\frac{ma^2}{\hbar^2} \frac{\partial E}{\partial a} =-8\pi\frac{m}{\hbar^2}\frac{\partial E}{\partial(1/a)}.
\end{equation}
This quantity determines how the energy changes as the scattering length changes and has a dimension of inverse length. Physically, it describes the short-range behavior of pairs of atoms (explicitly included in the H-LOCV method), or alternatively, the tail of the momentum distribution. In some cases, the intensive contact density $\mathcal{C}_2$, which has a dimension of $(length)^{-4}$, is used.  

Figure \ref{fig:Contact_N=1000} shows the two-body contact per particle $C_2/N$ for $N=1000$. It increases slowly at small $a$, then rises dramatically within the intermediate regime until it hits a maximum before saturating at unitarity. A similar behavior is also observed for different $N$, and in the renormalized Thomas-Fermi (TF) method \cite{Ding17_PRA} albeit in different units.

%DB \subsection{Limiting cases}
\subsection{Limiting Cases}

In the 
%DB in the 
H-LOCV model, once the value of $\nu$ is determined, properties of the condensate follow by solving the hyperradial equation for $F_{\nu}$.  Even simpler, to a good approximation the relative energy is given by the minimum value of the effective potential $V^{\rm eff}(\rho)$.   In two limiting cases,  $|a| \ll a_{ho}$ and $a \gg a_{ho}$, $\nu$ can be approximated analytically in the large-$N$ limit, hence so can the condensate's energy and contact.  These details are discussed in Appendices B and C.  Here, we explore the analytical results that follow, and compare these results to others in the literature.

\subsubsection{Small scattering length}

When $|a|$ is much smaller than the trap length $a_{ho}$, we find in Appendix \ref{appendix:WeakInteraction} that 
\begin{equation}
\label{eq:nuAtSmalla}
 \nu \underset{N\gg3}{\approx} \frac{1}{2}\sqrt{\frac{3}{\pi}} \frac{a}{\rho} N^{3/2}, \quad \text{if } \frac{|a|}{a_{ho}} \ll1.
\end{equation}
In this case the interaction potential takes the perturbative form 
\begin{align}
\label{eq:VintSmalla}
  V^{i\rm nt}(\rho) &= \frac{\hbar^2}{m \rho^2} \frac{3}{4}\sqrt{\frac{3}{\pi}}N^{7/2}\frac{a}{\rho}, \quad \text{for large $N$} .
 \end{align}
 This potential has the familiar $a/\rho^3$ scaling found already in the K-harmonic approximation and its extensions \cite{Bohn98_PRA,Sogo05_JPB}.  The size of the perturbation can be estimated by taking $\rho \approx \rho_0 \approx \sqrt{3N/2}\, a_{ho}$ at the minimum of the non-interacting effective potential (see Eq.~(\ref{eq:rho0})). Since $V^{\rm diag}(\rho_0)$ is of the order $N$, then $V^{\rm int}$ can be considered perturbative if $|a|/\rho \ll N^{-5/2}$. Then the total energy of the weakly interacting system can be approximated by
 \begin{equation}
E_0\approx \frac{3N}{2} \hbar \omega + \frac{1}{\sqrt{2\pi}} N^2 \frac{a}{a_{ho}} \hbar \omega. 
\end{equation}
A similar perturbative energy result emerges from the K-harmonic \cite{Bohn98_PRA} and hyperspherical-Faddeev \cite{Sogo05_JPB} methods. See inset of Fig.~\ref{fig:Contact_N=1000}(a).

\subsubsection{Infinite scattering length}

Of arguably greater interest is the resonant limit of infinite two-body scattering length, which is far from perturbative in many models.  However, the H-LOCV  method is capable of producing analytical results in this limit as well.  The details are worked out in Appendix C.

In this limit and at large $N$, the index $\nu$ has the form
\begin{align}
\label{eq:nu_infinity}
 \nu_{\infty}&= \left[ \frac{x_0}{\sqrt{6}}\left(\frac{27}{2\pi} \right)^{1/6}  \right]^2 N^{2/3}\\
 &\approx 2.122 \; N^{2/3}, \nonumber
 \end{align}
 where $x_0 \approx 2.798$ is the root to a transcendental equation defined in Appendix C.
 
 This asymptotic value of $\nu_{\infty}$ is independent of hyperradius in the $a \gg \rho$ limit, whereby it is easy to find the minimum of the effective potential.  This minimum is located at hyperradius
 \begin{eqnarray}
\label{eq:rho_infinity}
 \rho_\infty &=&  \left[ \frac{ x_0}{\sqrt{2}}\left(\frac{27}{2\pi} \right)^{1/6} \right]^{1/2} N^{2/3} a_{ho} \\
 &\approx& 1.588 \; N^{2/3} a_{ho}. \nonumber
\end{eqnarray}
From this, the condensate ground state energy is presented in Appendix C up to order $1/a$ (and ignoring the center of mass energy that is small in the large-$N$ limit)
\begin{align}
\label{eq:EnergyAtUnitarity}
 E_0 &\approx \sqrt{3N^2\nu_{\infty}}\left(1 -\frac{1}{2}\beta \left(\frac{\rho_\infty}{a}\right) \right) \hbar \omega \nonumber\\
 &\underset{N\rightarrow \infty}{\approx} \left( \frac{27}{16 \pi} \right)^{1/6} x_0 N^{4/3} \times \nonumber\\
 &\qquad\left[ 1 -  \left( \frac{16 \pi}{27} \right)^{1/12} N^{-1/6} \frac{x_0^2 + 1}{x_0^{7/2}} \frac{a_{ho}}{a}\right] \hbar \omega \\
 &
%DB = 
\approx
2.52 N^{4/3} \left( 1 - 0.254 N^{-1/6}\frac{a_{ho}}{a}\right) \hbar \omega,  \nonumber
\end{align}
where the function $\beta$ is defined in Eq.~(\ref{eq:beta}).
A consequence of this expansion is that we have an analytic expression for the contact of the resonant gas,
\begin{equation}
 C_2/N \approx 16.1 N^{1/6}/a_{ho}.
\end{equation}

The pair wave function used to construct the ground state wave function at unitarity has a node at $a^*=\sqrt{2}\rho_\infty\alpha_c$ where $\alpha_c$ is defined in Eq.~(\ref{eq:alphacAtUnitarity}), or 
\begin{align}
 a^{*} &\approx \frac{1}{\sqrt{x_0}}\frac{\pi}{2}\left(\frac{16\pi}{27}\right)^{1/12} N^{-1/6} a_{ho} \\ &\approx 0.989\, N^{-1/6} a_{ho}. \nonumber
\end{align}
This quantity serves as the effective scattering length and is about $30\%$ greater than the renormalized scattering length, $a^{*}_{TF}$, found in Ref. \cite{Ding17_PRA}:
\begin{equation}
 a^{*}_{TF}= \frac{2.182}{(6\pi^2)^{1/3}\langle n^{1/3} \rangle}, \quad \langle n^{1/3} \rangle \approx 0.4282 \frac{N^{1/6}}{a_{ho}}.
 \label{eq:astar}
\end{equation}

Further, in this limit the breathing mode frequency of the condensate  is easily derived.  It is simply the  oscillation frequency in the hyperradial potential, given by 
\begin{equation}
 \omega_b = \sqrt{\frac{1}{m} \left. \frac{\mathrm{d}^2 V_{eff}}{\mathrm{d} \rho^2} \right|_{\rho=\rho_{min}}}.
\end{equation}
On resonance, this quantity is given by 
\begin{equation}
 \omega_b = \omega \sqrt{ 6 \left( \frac{3N^2 \nu_\infty}{2}\right) \left(\frac{\rho_\infty}{a_{ho}}\right)^{-4}  +1} = 2\omega,
\end{equation}
a result already worked out long ago based on symmetry considerations \cite{Werner_PRA74,Castin_Physique5}.

The ground state of a resonant Bose gas has been considered previously for a homogeneous gas, as well as, more recently, for a trapped gas. Listed in Table~\ref{table:Energy} are the resulting energies from these previous calculations.  Because the H-LOCV method is intrinsically tied to a trapped gas, it is hard to compare directly to other calculations.  However, the result of Ref. \cite{Ding17_PRA} provides a link, by first calculating fixed-density quantities, then translating them into trapped values by means of the local-density approximation.  

In the homogeneous case, the energy per particle in the resonant limit $a \rightarrow \infty$ is expected to be a multiple of the characteristic (Fermi) energy $\hbar^2 n^{2/3} / 2m$ associated with the density $n$.  Several such values are reported in the table, spanning a range of about  a factor of four for the uniform system, and two for the trapped gas.  In the case of Ref. \cite{Ding17_PRA}, the renormalized scattering length affords a different energy at each value of density, whereby the energy can be represented in terms of the mean value $\hbar^2 \langle n^{2/3} \rangle / 2m$ averaged over an assumed Thomas-Fermi density profile.  Using this same density profile, one can write the energy per particle in terms of the characteristic energy scale of the trap, $\hbar \omega$, whereby a direct comparison can be made with the H-LOCV method.  This is done in the third column of Table I.   Just as the homogeneous LOCV appears to come in on the high side for the ground state energy, so too does  the H-LOCV method for the trapped gas.

\begin{table}[h!]
\caption{Ground state energy per particle, computed by various methods. The uniform gas energies are given in units of $\hbar^2 n^{2/3}/2m$, while the result of Ref.~\cite{Ding17_PRA} for a trapped system is given in units of  $\hbar^2 \langle n^{2/3}\rangle/2m$  with the mean density $\langle n^{2/3} \rangle$ determined by averaging over a Thomas-Fermi profile in a trap.  This same averaging allows this energy to be written in terms of the trap frequency $\omega$ in the final column.   }
\centering
\begin{center}
\begin{tabular}{|m{10em} |m{2cm}| m{2cm}|} 
%\begin{tabu} to 0.8\textwidth { | X[l] | X[c] | X[r] | }
 \hline 
 $E_0/N$ (uniform gas) & ($\hbar^2 n^{2/3}/2m$) &    \\ [0.5ex] 
 \hline \hline
 Cowell (LOCV) \cite{Cowell02_PRL} & 26.66 & -  \\
 \hline
  Song (Condensate amplitude variation) \cite{Song09_PRL} & 7.29 & - \\
 \hline
 Lee (Renormalization group) \cite{Lee10_PRA} & 6.0 &- \\
 \hline
 Diederix (Hypernetted chain) \cite{Diederix11_PRA} & 7.60 & - \\
 \hline
 Borzov (Resummation scheme) \cite{Borzov} & 8.48 & - \\
 \hline
 Zhou (RG) \cite{FZhou} & 8.11 & -\\
 \hline
 Yin (self-consistent Bogoliubov) \cite{Yin13_PRA} & 7.16 & - \\
 \hline
 van Heugten (renormalization group) \cite{vanHeugten} & 12.94 & -\\
 \hline
 Rossi (Monte-Carlo LOCV) \cite{Rossi_PRA89}  & 10.63 & -  \\
  \hline \hline
 $E_0/N$ (trapped gas) & $(\hbar^2 \langle n^{2/3}\rangle/2m)$ & $(N^{1/3} \hbar \omega)$   \\
 \hline \hline
 Ding (Renormalized K-harmonic) \cite{Ding17_PRA} & 12.67 & 1.205  \\
  \hline
 H-LOCV & - & 2.52  \\[1ex] 
 \hline
\end{tabular}
\end{center}
\label{table:Energy}
\end{table}

\begin{table}[h!]
\caption{Contact densities, computed by various methods. For the uniform gas, the intensive contact 
%DB $\mathcal{C}_2$ densities 
densities $\mathcal{C}_2$ 
are given in units of $n^{4/3}$, while the extensive contact $C_2/N$, in  units of $\langle n^{1/3} \rangle$, is given in Ref.~\cite{Ding17_PRA} for a trapped gas. Averaging over a Thomas-Fermi profile in a trap allows this $C_2/N$ to be written in terms of the characteristic trap length $a_{ho}$ in the final column.   }
\centering
\begin{center}
\begin{tabular}{|m{18em} |m{1.1cm}|m{1.5cm}|} 
%\begin{tabu} to 0.8\textwidth { | X[l] | X[c] | X[r] | }
 \hline 
 $\mathcal{C}_2$ (uniform gas) & $(n^{4/3})$ &    \\ [0.5ex] 
 \hline \hline
 Rossi (Monte-Carlo LOCV) \cite{Rossi_PRA89}  & 9.02 & -  \\
 \hline
  Diederix (Hypernetted chain) \cite{Diederix11_PRA} & 10.3 & - \\
 \hline
 Sykes \cite{Sykes14_PRA} & 12 & - \\
  \hline
 van Heugten (renormalization group) \cite{vanHeugten} & 32 & -\\
 \hline
 Yin (self-consistent Bogoliubov) \cite{Yin13_PRA} & 158 &- \\
 \hline 
 Smith (Virial theorem) \cite{Smith14_PRL} & 20 & - \\
\hline \hline
 $C_2/N$ (trapped gas) & $(\langle n ^{1/3} \rangle)$ & $(N^{1/6}/a_{ho})$   \\
 \hline \hline
 Ding (Renormalized K-harmonic) \cite{Ding17_PRA} & 11.8 & 5.05 \\
   \hline
 H-LOCV & - & 16.1  \\[1ex] 
 \hline
\end{tabular}
\end{center}
\label{table:Contact}
\end{table}

Likewise, there are various estimates for the contact on resonance, summarized in Table \ref{table:Contact}.  For a homogeneous system on resonance, one reports the intensive, density-dependent contact density $\mathcal{C}_2=\gamma n^{4/3}$, where $n$ is the gas density and $\gamma$ is a dimensionless constant.  This quantity, like the energy, is subject to an array of values tied to the different methods.  

For the trapped gas, fewer examples of the contact have been calculated.  We again turn to the work of Ref.~\cite{Ding17_PRA} to make the link.  For a trapped gas, one can describe an extensive contact $C_2$, related to the contact density by $C_2=\gamma N \langle n^{1/3}\rangle$, using the local density approximation (LDA). In this approximation, the trap results of 
%DB \cite{Ding17_PRA} 
Ref. \cite{Ding17_PRA} 
give $\gamma=11.8$ 
%DB (See 
(see
Table \ref{table:Contact}.).  Further, using the average value of $n^{1/3}$ in the Thomas-Fermi approximation, (\ref{eq:astar}), this result can be translated into natural harmonic oscillator units, yielding $C_2 = 5.05 N^{1/6}/a_{ho}$.  Compared to this, our value, also in the table, is $16.1 N^{1/6}/a_{ho}$.  Like the energy, the H-LOCV seems to overestimate the contact on resonance.  It does, however, correctly identify the $N^{1/6}$ number dependence of the contact for a trapped gas.

% Since $C_2/N$ is a density-dependent quantity, it should be obvious that in $a_{ho}^{-1}$ units, $C_2/N$ at unitarity will vary as a function of $N$. Using the energy expression in Eq.~(\ref{eq:EnergyAtUnitarity}), the LOCV contact is given by
%\begin{equation}
% C_2/N \approx 16.1 N^{1/6}/a_{ho}.
%\end{equation}
%Figure \ref{fig:ContactAtUnitarity} plots the $\sim N^{1/6}$ behaviour of $C_2/N$ at unitarity. The dashed line is the theoretical prediction of the renormalized TF method where $C_2/N=11.8 \langle n^{1/3} \rangle$, with $\gamma=11.8$ and $\langle n^{1/3} \rangle \approx 0.428 N^{1/6}/a_{ho}$ \cite{Ding17_PRA} which comes out to be  $C_2/N \approx 5.053 N^{1/6}/a_{ho}$, and smaller than the  LOCV result by about a factor of $1/3$. It is interesting to note that $a^{*}$ in \cite{Ding17_PRA} is larger than the LOCV $a^{*}$ by a third. A separate $\gamma$ prediction for a trapped system, also using LDA, gives $\gamma \approx 22$ \cite{Smith14_PRL}, which is about twice as large as the renormalized TF result hence a factor of $2/3$ smaller than the LOCV.

\section{Conclusions}

Despite its essential simplicity, the H-LOCV method provides a reasonable qualitative description of the mechanically stable Bose gas on resonance.  Notably, the method affords analytical estimates of essential quantities such as energy per particle and contact, when the scattering length is infinite.  It must be remembered that the approximation used here is only the lowest-order version of a hyperspherical theory that incorporates two-body correlations.  Various improvements can be made, including:

1) Extension to excited states.  We have so far incorporated only a single adiabatic channel function, consistent with our immediate goal of approximating a ground state.  Yet there exists a whole spectrum of states corresponding to different $\phi_{\nu}(\rho;\alpha)$.  We can contemplate states in which one or more particles are placed in excited states, corresponding to excitations of the BEC; or in the nodeless state {\it below} the condensate state, standing for bound molecular pairs.  We can also contemplate placing all the pairs in this nodeless state, to approximate the liquid-like configuration of Refs. \cite{Kalas_PRA,Gao05_PRL}.  In any case, having a spectrum of approximate energy eigenstates is a place to begin looking at the dynamics of a BEC quenched to resonance, or to any value of $a$.  Along with this, non-adiabatic couplings and their effect can be evaluated.

2) Extension of the Hamiltonian.  Thus far only two-body interactions have been contemplated, leading to pairwise Bethe-Peierls boundary conditions and universal behavior.  In more realistic treatments, additional three-body interactions are required and can also be incorporated.  In such a case, the hyperspherical basis set can be extended 
%DB I would suggest to take out the in a straightforward way---it might make it harder to publish your next paper on this...
%DB in a straightforward way 
to incorporate triplets of atoms, just as pairs were used here.  This involves adding an additional Jacobi coordinate and in principle several new hyperangles.  The machinery for this extension is well-known, yet incorporating it into the H-LOCV formalism requires careful attention. 

3) Extension of the Jastrow method.  A key approximation in the H-LOCV method has been to evaluate important integrals by approximating two-body correlation functions as in (\ref{eq:correlation_approximation}).  This rather severe approximation can be ameliorated, for example by a perturbation expansion known as the hypernetted chain approximation \cite{GreenBook}.  We are presently working to develop all these improvements.  

\begin{acknowledgements}
The authors thank J. P. Corson for lending us his code for the mean-field calculation and for careful reading of the manuscript, and J. P. D'Incao for discussions.

 M. W. C. S. and J. L. B. were supported by the JILA NSF Physics Frontier Center, grant number PHY-1734006, and by an ARO MURI Grant, number W911NF-12-1-0476.
%DB
D. B. was supported by the National Science Foundation through
grant numbers
PHY-1509892 and PHY-1745142.
 \end{acknowledgements}

\clearpage
\appendix

\section{Kinetic Energy Integrals}
\label{appendix:AngularKinetic}

To put the adiabatic basis functions to use, we must compute the matrix element that appears in Eq.~(\ref{eq:adiabatic_equations}).  In general this matrix element involves 
%DB complex 
complicated
multidimensional integrals.  However, far simpler, approximate versions of these integrals are often possible by means of the cluster expansion.  This is the same set of ideas developed originally in statistical mechanics
%DB , 
to derive virial coefficients in the not-quite-ideal gas equation of state. 

% Here we will adopt a simplified derivation of this kind of expansion, due to van Kampen \cite{vanKampen61_Physica}.  
We need to evaluate matrix elements of the grand angular momentum operator $\Lambda_{N-1}^2$.  Because our wave functions consider only one pair of atoms at a time, it should be sufficient to consider only the leading term $\Pi^2$, acting on the pair $(ij)=(12)$, and get the rest from symmetry.

To do so, let us for a moment return to independent-particle notation.  The kinetic energy $T$ is a sum of single-particle operators that acts on a pairwise-symmetrized wave function:
\begin{align}
\sum_{k=1}^{N} T_k \prod_{i<j}^{N} \phi_{\nu}(ij).
\end{align}
Choosing a single atom, say atom 1, the operator $T_1$ acts identically on $N-1$ of the terms in the product. Because $N$ atoms do the same, the action of  $T$ on the wave function can be written, for purposes of taking matrix elements, as (this is the substance of what Jastrow derives in his original paper \cite{Jastrow55_PR})
\begin{align}
N(N-1) [ T_1 \phi(12) ] \prod_{i<j, (ij) \ne (12)} \phi_{\nu}(ij),
\end{align}
where $T_1$ acts on a single pair, here chosen to be the pair $(12)$.  

Translated into the Jacobi coordinate $\eta_{N-1} = r_{12}/\sqrt{2}$, the kinetic energy acting on $\phi_{\nu}(12)$ is given by
\begin{align}
T_1 = - \frac{ \hbar ^2 }{ 2 m} \nabla_{1}^2 =- \frac{ \hbar ^2 }{ 2 m} \nabla_{12}^2 = - \frac{ \hbar ^2 }{ 2 m } 
\left( \frac{ 1 }{ 2 } \nabla_{\eta_{N-1}}^2  \right)
= \frac{ \hbar ^2 }{ 2m } \frac{ 1 }{ 2 \rho^2 } \Pi^2, 
\end{align}
ignoring the $\rho$-derivatives that are set to zero in the adiabatic approximation.
Moreover, $\phi_{\nu}$ is an eigenfunction of $\Pi^2$ in the interval $0 \le \alpha \le \alpha_d$, as described in (\ref{eq:evalue}). Beyond $\alpha_d$, $\phi_\nu(\alpha)$ is constant and $\Pi^2 \phi_\nu(\alpha)$ vanishes (where the operator $\Pi^2$ is defined in (\ref{eq:Pi2_definition})). Thus, the action of the hyperangular kinetic energy becomes
\begin{widetext}
\begin{align}
\Lambda_{N-1}^2 \prod_{i<j} \phi_{\nu}(ij) = 
\left\{ \begin{array}{ll}
\frac{ 1 }{ 2 } N(N-1) 2\nu(2\nu+3N-5) \prod_{i<j} \phi_{\nu}(ij), & 0 \le \alpha \le \alpha_d \\
0, & \alpha > \alpha_d
\end{array} \right.  
%DB 
.
\end{align}
\end{widetext}
For a single channel calculation, the expectation value of the grand angular momentum is, therefore, given by
%DB please check the change I made...
\begin{widetext}
\begin{align}
\langle \nu | \Lambda_{N-1}^2 | \nu \rangle
&= \frac{ 1 }{ 2 } N(N-1) 2\nu(2\nu+3N-5)
\frac{ 4 \pi \int_0^{\alpha_d} d \Omega_{\alpha} \int d\Omega_{N-2} 
\prod_{i<j} |\phi_{\nu}(ij)|^2 }
{ \int d\Omega_{N-1} \prod_{i<j} 
%DB \phi^2_{\nu}(ij) 
|\phi_{\nu}(ij)|^2 
} \nonumber\\
&=  \frac{ N }{ 2 } 2\nu(2\nu+3N-5),
\end{align}
\end{widetext}
where the last equality is obtained by enforcing the LOCV normalization condition that sets Eq.~(\ref{eq:LOCV_integral}) to unity.

%DB \section{Weak interaction in the large $N$ limit}
\section{Weak Interaction in the Large $N$ Limit}
\label{appendix:WeakInteraction}
The ground state solution for 
%DB
the
non-interacting system, $a=0$, gives $\nu=0$, $B=0$, and 
\begin{equation}
 \phi_{\nu}(\alpha_d)=A = 1.
\end{equation}
The LOCV boundary condition (\ref{eq:BC_norm}) becomes
\begin{equation}
\label{eq:LOCVbcLargeN}
1 \approx \frac{4}{\sqrt{\pi}} \left( \frac{3}{2} \right)^{3/2} N^{5/2} \int_0^{\alpha_d} \mathrm{d}\alpha\, \alpha^2 A^2 = 3\sqrt{\frac{6}{\pi}} N^{5/2} \frac{\alpha_d^3}{3},
 \end{equation}
 where we used
 \begin{equation}
  \frac{ \Gamma( \frac{3N-3}{2} ) }{ \Gamma( \frac{3N-6}{2}) } \approx \frac{\left( \frac{3N-3}{2} \right)^{ (3N-4)/2 }}{\left( \frac{3N-6}{2} \right)^{ (3N-7)/2 }} \approx \left(\frac{3N}{2} \right)^{3/2}, \quad N\rightarrow \infty.
 \end{equation}
Thus,
\begin{equation}
\label{eq:alphadZeroa}
 \alpha_d \approx (\pi/6)^{1/6} N^{-5/6}, \quad \text{if } a=0,
\end{equation}
which is an extremely small quantity when $N \rightarrow \infty$. We assume the same scaling for $\alpha_d$ when $0<a\ll a_{ho}$. Now, for small nonzero $a$, $B \neq 0$, boundary condition (\ref{eq:BC_alphad}) becomes
%DB I changed to normal round brackets to left/right brackets...
\begin{align}
  %\frac{B}{A} &= -\frac{a}{\sqrt{2}\rho}, \nonumber\\
  0 &= \left. \frac{ \partial \phi_{\nu} }{ \partial \alpha } \right\vert_{\alpha_d} \approx A f_\nu^{\prime} (\alpha_d) + B g_\nu^{\prime} (\alpha_d), \nonumber\\
  f_\nu^{\prime} (\alpha) & \underset{\substack{\sqrt{3N\nu}\alpha \ll 1,\\ \alpha \ll 1}}{\approx} -\frac{4}{3} \nu\left(\frac{3N-5 }{ 2 } + \nu \right)\alpha, \\
  g_\nu^{\prime} (\alpha) & \underset{\substack{\sqrt{3N\nu}\alpha \ll 1,\\ \alpha \ll 1}}{\approx} -\frac{1}{\alpha^2} + 4
%DB (-\nu -\frac{1}{2}) 
\left(-\nu -\frac{1}{2} \right) 
\left(\frac{3N-6 }{ 2 } + \nu \right) ,
  \end{align}
 which gives
 \begin{align}
  \frac{A}{B} \underset{N\rightarrow \infty}{\approx} -\sqrt{\frac{3}{2 \pi }}\frac{1}{\nu} N^{3/2}.
  \end{align}
With the Bethe-Peierls boundary condition~(\ref{eq:BC_alpha0}), we get
\begin{equation}
\label{eq:nuAtSmalla}
 \nu \approx \frac{1}{2}\sqrt{\frac{3}{\pi}} \frac{a}{\rho} N^{3/2}, \quad \text{if } \frac{|a|}{a_{ho}} \ll1.
\end{equation}
Note that $\nu$ is not an absolutely small quantity as it also depends on $N$.

%DB \section{Strong interaction in large $N$ limit}
\section{Strong Interaction in the Large $N$ Limit}
\label{appendix:StrongInteraction}
As $a \rightarrow +\infty$, we assume from Eq.~(\ref{eq:BC_alpha0}) that $A \approx 0$ and
\begin{align}
\label{eq:phiAtUnitarity}
 \phi_{\nu}(\alpha) &\underset{N>\nu_\infty \gg \frac{1}{2}}{\approx} B \alpha^{-1} {}_2F_1 \left( - \nu_\infty, \frac{ 3N}{ 2 }, \frac{ 1 }{ 2 }; \alpha^2 \right) \nonumber\\
 &=B \frac{\cos{\sqrt{6N\nu_\infty\alpha^2}}}{\alpha},
\end{align}
where we used
\begin{align}
 {}_2F_1 \left( - \nu, \frac{ 3N}{ 2 }, \frac{ 1 }{ 2 }; \alpha^2 \right) &= \sum_{k=0}^\infty \frac{(-1)^k}{(2k-1)!! k!} (3N\nu\alpha^2)^k\nonumber\\
 &= \cos{\sqrt{6N\nu\alpha^2}}.
 \end{align}
 Note that $\alpha$ may be small but the product $\sqrt{6N\nu_\infty}\alpha$ need not be. Also, this wave function is zero when $\alpha=\alpha_c$:
 \begin{equation}
 \label{eq:alphacAtUnitarity}
  \alpha_c =\frac{\pi}{2}\frac{1}{\sqrt{6N\nu_\infty}} .
 \end{equation}
 Boundary conditions~(\ref{eq:BC_alphad}) and~(\ref{eq:BC_norm}) yield the relations
 \begin{align}
 \label{eq:B}
   B &= \frac{\alpha_d}{\cos{\sqrt{6N\nu_\infty\alpha_d^2}}}\\
  \label{eq:transcendental}
   0 &=1 + \sqrt{6N\nu_\infty\alpha_d^2} \tan \sqrt{6N\nu_\infty\alpha_d^2} \\
  \label{eq:LOCVbcUnitarity}
   1 &\approx \frac{4}{\sqrt{\pi}} \left( \frac{3}{2} \right)^{3/2}  \frac{N^{5/2}\alpha_d^3}{\cos^2 {\sqrt{6N\nu_\infty\alpha_d^2}}} \frac{1}{2} \left[1  + \frac{\sin\left({2\sqrt{6N\nu_\infty\alpha_d^2}}\right)}{2\sqrt{6N\nu_\infty\alpha_d^2}}  \right].
 \end{align}
From Eqs.~(\ref{eq:transcendental}) and (\ref{eq:LOCVbcUnitarity}), we get
 \begin{equation}
 \label{eq:alphadUnitarity}
  \alpha_d=\left(\frac{2\pi}{27} \right)^{1/6} N^{-5/6} \approx 0.7843\, N^{-5/6}, \quad \text{if } a \rightarrow +\infty.
 \end{equation}
 Equation ~(\ref{eq:transcendental}) has the form
 \begin{equation}
  1+x_0 \tan x_0=0,
 \end{equation}
with solutions  $x_0 \approx 2.798,\, 6.121,...$. If $x_0=\sqrt{6N\nu_\infty\alpha_d^2}$, then 
\begin{align}
\label{eq:nuAtUnitarity}
 \nu_{\infty}&= \left[ \frac{x_0}{\sqrt{6}}\left(\frac{27}{2\pi} \right)^{1/6}  \right]^2 N^{2/3}\\
 &\approx 2.122\, N^{2/3} \quad\text{if } x_0=2.798. \nonumber
 \end{align}
 
 Now, if $0<a_{ho}/a\ll1$, then
 \begin{align}
\label{eq:phiAtBiga}
 \phi_{\nu}(\alpha) &\underset{N>\nu \gg \frac{1}{2}}{\approx}A\, {}_2F_1 \left( - \nu, \frac{ 3N}{ 2 }, \frac{ 3 }{ 2 }; \alpha^2 \right) + \nonumber\\
 &\quad\qquad\qquad B \alpha^{-1} {}_2F_1 \left( - \nu, \frac{ 3N}{ 2 }, \frac{ 1 }{ 2 }; \alpha^2 \right) \nonumber\\
 &\quad\approx A \frac{\sin{\sqrt{6N\nu\alpha^2}}}{\sqrt{6N\nu\alpha^2}} + B \frac{\cos{\sqrt{6N\nu\alpha^2}}}{\alpha}.
\end{align}
Boundary condition~(\ref{eq:BC_alphad}) yields 
\begin{equation}
 \frac{A}{B} = \frac{1 + \sqrt{6N\nu\alpha_d^2} \tan \sqrt{6N\nu\alpha_d^2}}{\sqrt{6N\nu\alpha_d^2} -\tan \sqrt{6N\nu\alpha_d^2}}\sqrt{6N\nu}
\end{equation}
With the Bethe-Peierls boundary condition~(\ref{eq:BC_alpha0}), we get the relation
\begin{align}
\label{eq:transcendentalBiga}
 \frac{1 + x \tan x}{x -\tan x} &= - \epsilon, \quad \text{ where}\\
 x&=\sqrt{6N\nu\alpha_d^2},\\
 \epsilon &=  \frac{\rho}{\sqrt{3N\nu}}\frac{1}{a}.
\end{align}
Let $x=x_0 - \Delta$. If $\epsilon=0$, then we recover Eq.~(\ref{eq:transcendental}) and $\Delta=0$. Suppose $0<\epsilon \ll 1$ so that $\Delta$ is also a small varying quantity. Then, using 
\begin{equation}
 \tan x \approx \frac{\tan x_0 -\Delta}{ 1 + \Delta \tan x_0}, \nonumber
 \end{equation}
Eq.~(\ref{eq:transcendentalBiga}) gives
 \begin{equation}
  \Delta \approx \frac{x_0^2 +1}{ x_0^2} \epsilon.
 \end{equation}
 Expressing $x$ and $\epsilon$  back in terms of $\nu$ and $a$, 
 \begin{equation}
 \label{eq:nuVsOneOna}
  \sqrt{6N\nu} \alpha_d = x_0 -\frac{x_0^2 +1}{ x_0^2} \frac{\rho}{\sqrt{3N\nu}}\frac{1}{a}.
 \end{equation}
 Note that if the second term vanishes, then $\nu=\nu_\infty$ as defined in Eq.~(\ref{eq:nuAtUnitarity}). Let $\nu=\nu_\infty(1-\beta)$, where $\beta$ is a function of $1/a$. Then Eq.~(\ref{eq:nuVsOneOna}) leads to
 \begin{align}
   &\sqrt{6N\nu_{\infty}}\left(1-\frac{1}{2}\beta \right) \alpha_d \approx x_0 -\frac{x_0^2 +1}{ x_0^2} \frac{\rho}{\sqrt{3N\nu_\infty}}\frac{1}{a} \left(1+\frac{1}{2}\beta \right) \nonumber\\
   \label{eq:beta}
  \quad & \Rightarrow \quad \beta \approx 2\frac{x_0^2 +1}{ x_0^3}\frac{\rho}{\sqrt{3N\nu_\infty}}\frac{1}{a}.
 \end{align}


\begin{thebibliography}{20}

\bibitem{Makotyn14_NP} 
P. Makotyn, C. E. Klauss, D. L. Goldberger, E. A. Cornell, and D. S. Jin, Nat. Phys. {\bf 10}, 116 (2014).

\bibitem{Klauss17_PRL} C. E. Klauss, X. Xie, C. Lopez-Abadia, J. P. D'Incao, Z. Hadzibabic, D. S. Jin, and E. A. Cornell, Phys. Rev. Lett. {\bf 119}, 143401 (2017).

\bibitem{Eigen17_PRL} C. Eigen, J. A. P. Glidden, R. Lopes, N. Navon, Z. Hadzibabic, and R. P. Smith,  Phys. Rev. Lett. {\bf 119}, 250404 (2017).

\bibitem{Fletcher17_Science} R. J. Fletcher, R. Lopes, J. Man, N. Navon, R. P. Smith, M. W. Zwierlein, and Z. Hadzibabic, Science {\bf 355}, 377 (2017).

\bibitem{Fletcher13_PRL} R. J. Fletcher, A. L. Gaunt, N. Navon, R. P. Smith and Z. Hadzibabic, Phys. Rev. Lett. {\bf 111},  125303 (2013).

\bibitem{Song09_PRL} 
J. L. Song and F. Zhou, Phys. Rev. Lett. {\bf 103}, 025302 (2009).

\bibitem{Yin13_PRA} X. Yin and L. Radzihovsky, Phys. Rev. A {\bf 88}, 0636121 (2013).

\bibitem{Sykes14_PRA} A. G. Sykes, J. P. Corson, J. P. D'Incao, A. P. Koller, C. H. Greene, A. M. Rey, K. R. A. Hazzard, and J. L. Bohn, Phys. Rev. A {\bf 89}, 021601 (2014).

\bibitem{Yin16_PRA} X. Yin and L. Radzihovski, Phys. Rev. A {\bf 93}, 033653 (2016).

\bibitem{Corson15_PRA} J. P. Corson and J. L. Bohn, Phys. Rev. A {\bf 91}, 013616 (2015).

\bibitem{Corson16_PRA} J. P. Corson and J. L. Bohn, Phys. Rev. A {\bf 94}, 023604 (2016).

\bibitem{Ding17_PRA} Y. Ding and C. H. Greene, Phys. Rev. A {\bf 95}, 053602 (2017).

\bibitem{vanHeugten} H. T. C. Stoof and J. J. R. M. van Heugten, J. Low Temp. Phys. {\bf 174}, 159 (2014).

\bibitem{Bohn98_PRA} J. L. Bohn, B. D. Esry, and C. H. Greene, Phys. Rev. A {\bf 58}, 584 (1998).

\bibitem{Das04_PRA} T. K. Das and B. Chakrabarti, Phys. Rev. A {\bf 70}, 063601 (2004).

\bibitem{Das07_PRA} T. K. Das, S. Canuto, A. Kundu, and B. Chakrabarti, Phys. Rev. A {\bf 75}, 042705 (2007). 

\bibitem{Chakrabarti08_PRA} B. Chakrabarti and T. K. Das, Phys. Rev. A {\bf 78}, 063608 (2008).

\bibitem{Lekala14_PRA} M. L. Lekala, B. Chakrabarti, G. J. Rampho, T. K. Das, S. A. Sofianos, and R. M. Adam, Phys. Rev. A {\bf 89}, 023624 (2014).

\bibitem{Sorensen02_PRA} O. S{\o}rensen, D. V. Fedorov, and A. S. Jensen, Phys. Rev. {\bf A} 66, 032507 (2002).  

\bibitem{Sorensen03_PRA} O. S{\o}rensen, D. V. Fedorov, and A. S. Jensen, Phys. Rev. A {\bf 68}, 063618 (2003).  

\bibitem{Sorensen04_JPB} O. S{\o}rensen, D. V. Fedorov, and A. S. Jensen, J. Phys. B {\bf 37}, 93 (2004).

\bibitem{Sogo05_JPB} T. Sogo, O. S{\o}rensen, A. S. Jensen, and D. V. Fedorov, J. Phys. B {\bf 38}, 1051 (2005).

\bibitem{Jastrow55_PR} R. Jastrow, Phys. Rev. {\bf 98}, 1479 (1955).

\bibitem{Pandharipande73_PRC} V. R. Pandharipande and H. A. Bethe, Phys. Rev. C {\bf 7}, 1312 (1973).  

\bibitem{Pandharipande77_PRA} V. R. Pandharipande and K. E. Schmidt, Phys. Rev. A {\bf 15}, 2486 (1977).

\bibitem{Cowell02_PRL} S. Cowell, H. Heiselberg, J. Morales, V. R. Pandharipande, and C. J. Pethick, Phys. Rev. Lett. {\bf 88}, 210403 (2002).

\bibitem{Smirnov77} Yu. F. Smirnov and K. V. Shitikova, Sov. J. Part. Nucl. {\bf 8}, 344 (1977).

\bibitem{Avery97_IJQC} J. Avery, W. Bian, J. Loeser, and F. Antonsen, Int. J. Quant. Chem. {\bf 63}, 5 (1997).

\bibitem{Aquilanti86_JCP} V. Aquilanti, S. Cavalli, and G. Grossi, J. Chem. Phys. {\bf 85}, 1362 (1986).

\bibitem{Avery_book} J. Avery, {\it Hyperspherical Harmonics and Generalized Sturmians} (Kluwer, New York, 2002).

\bibitem{Blume00_JCP} D. Blume, C. H. Greene, and B. D. Esry, J. Chem. Phys. {\bf 113}, 2145 (2000).

\bibitem{Avery_book2} J. Avery, {\it Hyperspherical Harmonics: Applications in Quantum Theory} (Kluwer, Dordrecht, 1989).

\bibitem{Frey85_ChemPhys} J. G. Frey and B. J. Howard, Chemical Physics. {\bf 99}, 415 (1985).

\bibitem{Starace79_PRA} A. F. Starace and G. L. Webster, Phys. Rev. A {\bf 19}, 1629 (1979).

\bibitem{Werner_PRA74} F. Werner and Y. Castin, Phys. Rev. A {\bf 74}, 053604 (2006).

\bibitem{GreenBook} A. Fabrocini, S. Fantoni, and E. Krotscheck, {\it Ed.} {\it Introduction to Modern Methods of Quantum Many-Body Theory and Their Applications} (World Scientific, 2002), Ch. 2.

\bibitem{Erdelye53} A. Erd\'ele, {\it Ed.} {\it Higher Transcendental Functions, Vol II} (McGraw-Hill, 1953), p. 169. 

\bibitem{Kalas_PRA} R. M. Kalas and D. Blume, Phys. Rev. A {\bf 76}, 013617 (2007).

\bibitem{Rossi_PRA89} M. Rossi, L. Salasnich,  F. Ancilotto and F. Toigo, Phys. Rev. A {\bf 89}, 041602(R) (2014).

\bibitem{Blume_preprint} D. Blume, M. W. C. Sze, and J. L. Bohn,  unpublished.

\bibitem{Gao05_PRL} B. Gao, Phys. Rev. Lett. {\bf 95}, 240403 (2005).

\bibitem{Jonsell02_PRL} S. Jonsell, H. Heiselberg and C. J. Pethick, Phys. Rev. Lett. {\bf 89}, 250401 (2002).

\bibitem{Braaten06_PhysRep} E. Braaten and H.-W. Hammer, Physics Reports {\bf 428}, 259 (2006).


\bibitem{Pethick} C. J. Pethick and H. Smith, {\it Bose-Einstein Condensation in Dilute Gases} (Cambridge University Press, Cambridge, 2002). 

% \bibitem{Dalfovo99_RevModPhys} F. Dalfovo, S. Giorgini, L. Pitaevskii, and S. Stringari, Rev. Mod. Phys. {\bf 71}, 463 (1999).

\bibitem{Blume01_PRA} D. Blume and C. H. Greene, Phys. Rev. A {\bf 63}, 063601 (2001).

\bibitem{Smith14_PRL} D. Smith, E. Braaten, D. Kang and L. Platter, Phys. Rev. Lett.  {\bf 112}, 110402 (2014).

\bibitem{Castin_Physique5} Y. Castin, C. R. Physique {\bf 5}, 407 (2004).

\bibitem{Lee10_PRA} Y. L. Lee and Y. W. Lee, Phys. Rev. A  {\bf 81}, 063613 (2010).

\bibitem{Borzov} D. Borzov, M. S. Mashayekhi, S. Zhang, J.-L. Song and
F. Zhou, Phys. Rev. A  {\bf 85}, 023620 (2012).

\bibitem{Diederix11_PRA} J. M. Diederix, T. C. F. van Heijst, and H. T. C. Stoof, Phys. Rev. A  {\bf 84}, 033618 (2011).

\bibitem{FZhou} F. Zhou and M. S. Mashayekhi, Ann. Phys. {\bf 328}, 83 (2013).





%\bibitem{vanKampen61_Physica} N. G. van Kampen, Physica {\bf 27}, 783 (1961).  

\end{thebibliography}
\end{document}